\documentclass[english,aps, prb,amsmath,amssymb,
twocolumn,
nofootinbib,superscriptaddress]{revtex4}
\usepackage[T1]{fontenc}
\usepackage[latin9]{inputenc}
\usepackage{color}
\usepackage{float}
\usepackage{amsbsy}
\usepackage{amstext}
\usepackage{graphicx}
\usepackage{esint}
\usepackage{epstopdf}
\usepackage{hyperref}

\usepackage{amsmath}

\usepackage[normalem]{ulem}

\makeatletter

\@ifundefined{textcolor}{}
{%
 \definecolor{BLACK}{gray}{0}
 \definecolor{WHITE}{gray}{1}
 \definecolor{RED}{rgb}{1,0,0}
 \definecolor{GREEN}{rgb}{0,1,0}
 \definecolor{BLUE}{rgb}{0,0,1}
 \definecolor{CYAN}{cmyk}{1,0,0,0}
 \definecolor{MAGENTA}{cmyk}{0,1,0,0}
 \definecolor{YELLOW}{cmyk}{0,0,1,0}
 }

\usepackage{dcolumn}
\usepackage{bm}
\usepackage{array}

\def\ket{\rangle}
\def\bra{\langle}

\def\HPT{H_\textrm{PT}}
\def\HPTop{\hat{H}_\textrm{PT}}

\def\gRIm{\bar{\gamma}_{\textrm{I}-}}
\def\gRIp{\bar{\gamma}_{\textrm{I}+}}
\def\gRImp{\bar{\gamma}_{\textrm{I}\mp}}
\def\gRII{\bar{\gamma}_\textrm{II}}
\def\gPT{\bar{\gamma}_\textrm{PT}}
\def\ggap{g_{\textrm{gap},1}}

\def\gEPf{\bar{\gamma}_\textrm{gap}}
\def\gEPfpm{\bar{\gamma}_{\textrm{gap},\pm}}
\def\gEPfp{\bar{\gamma}_{\textrm{gap},+}}
\def\gEPfm{\bar{\gamma}_{\textrm{gap},-}}

\def\ggapm{g_{\textrm{gap},2}}

\newcommand{\PT}{\mathcal{PT}}

\newcommand{\beq}{\begin{equation}}
\newcommand{\eeq}{\end{equation}}
\newcommand{\beqa}{\begin{eqnarray}}
\newcommand{\eeqa}{\end{eqnarray}}

\makeatother

\usepackage{babel}

\begin{document}

\title{Reservoir-assisted symmetry breaking and coalesced zero-energy modes in an open $\PT$-symmetric Su-Schrieffer-Heeger model}
\author{Savannah Garmon}
%
\affiliation{Department of Physical Science, 
Osaka Prefecture University, 
Gakuen-cho 1-1, Sakai 599-8531, Japan}
\affiliation{Institute of Industrial Science, University of Tokyo, Kashiwa 277-8574, Japan}
\author{Kenichi Noba}
\affiliation{Department of Physical Science, 
Osaka Prefecture University, 
Gakuen-cho 1-1, Sakai 599-8531, Japan}

\begin{abstract}
We study a model consisting of a central $\PT$-symmetric trimer with non-Hermitian strength parameter $\gamma$ coupled to two semi-infinite Su-Schrieffer-Heeger (SSH) leads.  
We show the existence of two zero-energy modes, one of which is localized while the other is anti-localized.
For the remaining eigenvalues, we demonstrate two qualitatively distinct types of $\PT$-symmetry breaking.  Within a subset of the parameter space corresponding to the topologically non-trivial phase of the SSH chains, a gap opens within the broken $\PT$ regime of the discrete eigenvalue spectrum.  For relatively smaller values of $\gamma$, the eigenvalues are embedded in the two SSH bands and hence become destabilized primarily due to the resonance interaction with the continuum.  We refer to this as reservoir-assisted $\PT$-symmetry breaking.  As the value of $\gamma$ is increased, the eigenvalues exit the SSH bands and the discrete eigenstates become more strongly localized in the central trimer region.  This approximate decoupling results in the discrete spectrum behaving more like the independent trimer, including both a region in which the $\PT$-symmetry is restored (the gap) and a second region in which it is broken again.  At the exceptional point (EP) marking the boundary between the gap and the second $\PT$-broken region, two of the eigenstates coalesce with the localized zero-energy mode, resulting in a third-order exceptional point (EP3).  At the other boundaries of the parameter space at which the gap vanishes, similar higher-order EPs can form as pairs of the discrete eigenstates coalesce with either of the two zero-energy states.  The EPs of order $N$ formed of the localized zero-energy state give rise to characteristic dynamics $\sim t^{2N-2}$ in the evolution of an initial state, which we propose to measure in a photonic lattice experiment. 
\end{abstract}

\maketitle

\section{introduction}
\label{sec:intro}

While the traditional formulation of quantum mechanics requires that the Hamiltonian operator describing a given system must be Hermitian in order to yield real eigenvalues, almost from the beginning of quantum mechanics researchers have found it useful to consider non-Hermitian extensions
or interpretations of the theory.  
Under one common approach, a non-Hermitian formulation is often useful to describe the interaction between a quantum system and its surrounding environment.  In this picture, the appearance of a resonance with complex eigenvalue is often associated with exponential decay.\cite{Gamow28,Siegert39,Nakanishi,Sudarshan78,PPT91,HSNP08,Rotter_review,Moiseyev,Madrid12,GO17,OH17A}

The observation by Bender and Boettcher in 1998 that non-Hermitian systems obeying parity time ($\mathcal{PT}$) symmetry can still yield real eigenvalues \cite{BB98} has lead some researchers to consider 
$\mathcal{PT}$-symmetric or pseudo-Hermitian reformulations of quantum mechanics \cite{Bender_review,BQZ01,MostaJMP02,BBJ02,MostaPRL07}.  
These efforts have in turn inspired studies of $\mathcal{PT}$-symmetry in a wide range of physical contexts, particularly optics \cite{RDM05,KGM08,Zheng10,MECM08,Christo12,Feng12,MostaPRA13,Peng14,Konotop_review}, but also electronic circuits \cite{PTCircuitExpt}, random walks \cite{Obuse16,PTQWexpt} and descriptions of open quantum systems \cite{GGH15,Shobe2021}.

A key issue that underlies many interesting properties in such systems is spontaneous $\PT$-symmetry breaking, under which at least two solutions of the $\PT$-symmetric Hamiltonian no longer conform to the $\PT$-symmetry individually but only do so as a pair.  The threshold of the $\PT$-symmetry breaking transition occurs at an exceptional point (EP), at which two or more eigenstates coalesce and the usual diagonalization scheme breaks down \cite{Kato,GraefeEP3,KGTP17,BerryEP,HeissEP,MiriAluEP}.
Owing to the broken $\PT$-symmetry, the physics on either side of the EP can be quite different.
As a simple example (that will be useful for our later development), consider a $3 \times 3$ matrix Hamiltonian of the form
\beq
  H_\textrm{PT} = 
  	\left( \begin{array}{ccc}
		i \gamma		& g	& 0	\\
		g			& 0	& g	\\
		0			& g 	& - i \gamma
	\end{array} \right)
	.
\label{HPT.matrix}
\eeq
This satisfies the $\PT$-symmetric relation $\HPT = \PT \HPT \PT$, in which the parity operator is defined as
\beq
  \mathcal{P}
  	= \left( \begin{array}{ccc}
		0	& 0	& 1	\\
		0	& 1 	& 0	\\
		1	& 0	& 0
		\end{array} \right)
	,
\eeq
while $\mathcal{T}$ is the anti-linear complex conjugation operator acting as $\mathcal{T} i \mathcal{T} = -i$  \cite{Bender_review}.  
Physically, we can interpret the non-Hermitian entries $\pm i \gamma$ of $\HPT$ as representing energy source and drain terms, respectively \cite{RDM05}.
The eigenvalues of $\HPT$ are given by $z_0 = 0$ and 
\beq
  z_\pm = \pm \sqrt{2g^2 - \gamma^2},
\label{HPT.solns}
\eeq
the latter two of which demonstrate the $\PT$-symmetric properties of the system.  The $\PT$-symmetric phase of $\HPT$ is given by $\gamma < \sqrt{2} g$, during which the effective coupling $\sqrt{2}g$ between the energy gain and loss terms 
is strong enough to balance their individual non-Hermitian character, resulting in real eigenvalues $z_\pm$.  Meanwhile, in the case $\gamma > \sqrt{2}g$, the non-Hermiticity overwhelms the coupling such that the eigenvalues $z_\pm$ become a complex conjugate pair; this is the broken $\PT$-symmetry regime.  
Of course, it is the EP at $\gamma = \gPT \equiv \sqrt{2}g$ that separates the two regions.

The above gives a relatively simple picture of $\PT$-symmetry breaking in a finite, purely discrete system.
However, the physics becomes significantly more complex when we combine $\PT$ symmetry in the form of gain and loss with traditional open quantum systems that incorporate both discrete and continuous spectra \cite{GGH15}.  
The continuum in such systems arises from microscopic degrees of freedom that describe the environment as it influences the quantized part of the system \cite{LNNB00,LonghiPRA06,PPT91,HSNP08,GGH15,GNHP09,HO14,GNOS19,KH11,Hatano13,GTC2D2,TGKP16,GOH21}.  
We note that in cavity quantum electrodynamics, a model incorporating such a continuum is sometimes referred to as a {\it structured reservoir} \cite{LNNB00,LonghiPRA06,GTC2D2,GOH21,PRX2}.

The interaction between the reservoir (continuum) and the gain/loss profile in the combined system can result in a wide variety of phenomena that cannot be accommodated in purely discrete $\PT$ models \cite{RDM05,MECM08,GGH15,Hamid14,LonghiLABS,KZ17}.  
As we show in this work, new physics particularly emerges when the energy of these two subsystems are roughly balanced.  In particular, when the energy scale of a $\PT$-symmetric defect and the continuum are similar, we find that the $\PT$-symmetry-breaking threshold occurs for significantly reduced values of 
the strength of the complex potential.  We further show that the eigenstates associated with the broken $\PT$ symmetry in this scenario take on properties of both the reservoir and the $\PT$-symmetric subsystem.
Hence, we refer to this as {\it reservoir-assisted $\PT$-symmetry breaking}.  
A special case occurs in the reservoir-assisted $\PT$-broken regime when one of the complex modes associated with the explicitly non-Hermitian sector of the model becomes embedded directly in the continuum.  This is known in the literature as a resonance-in-continuum (RIC) \cite{GGH15,Shobe2021} or spectral singularity \cite{Mostafa2009lett,Mostafa2009A,Mostafa2011,Longhi2009,Longhi10,ZK20}; physically, this can be understood as a coherent, non-equilibrium steady state in which particle flux from the gain/loss sector is reprocessed through the reservoir to form a standing wave extending into the surrounding environment.
In $\PT$-symmetric models these standing waves constitute simultaneous laser-absorber modes \cite{LonghiLABS,KZ17}.

In Sec. \ref{sec.Hamiltonian} we present our model, consisting of a $\PT$-symmetric central potential that is equivalent to the Hamiltonian from Eq. (\ref{HPT.matrix}), which is then coupled to the reservoir in the form of 
two semi-infinite Su-Schrieffer-Heeger 
(SSH) chains \cite{SSH1979,Asboth2016,BPCG19,PRX1}.
This model is a two-channel extension of the $\PT$-symmetric open quantum system
from our previous paper \cite{GGH15}, which is useful for distinguishing between the qualitatively different types of $\PT$-symmetry breaking.

Recently, non-Hermitian extensions of the SSH model have been actively investigated
by many authors who mainly focused on the topological properties of the 
systems.\cite{Zhu2014,Klett2017,Dangel2018,Jin2017,Yuce2018,Lieu2018,Yao2018,Kunst2018,Kong2020,HH20,Roccati21,PLA1,PLA2} 
In our model, the semi-infinite SSH chains form the reservoir.
We mainly focus on the parameter region corresponding to the topologically non-trivial phase of the bare SSH chains,\cite{Asboth2016} in which edge states can form that are topologically protected.
In our model, these can result in states that approximately decouple from the chains and are hence localized 
around the central $\PT$-symmetric impurity region.

\begin{figure*}[t]
  \includegraphics[width=12cm]{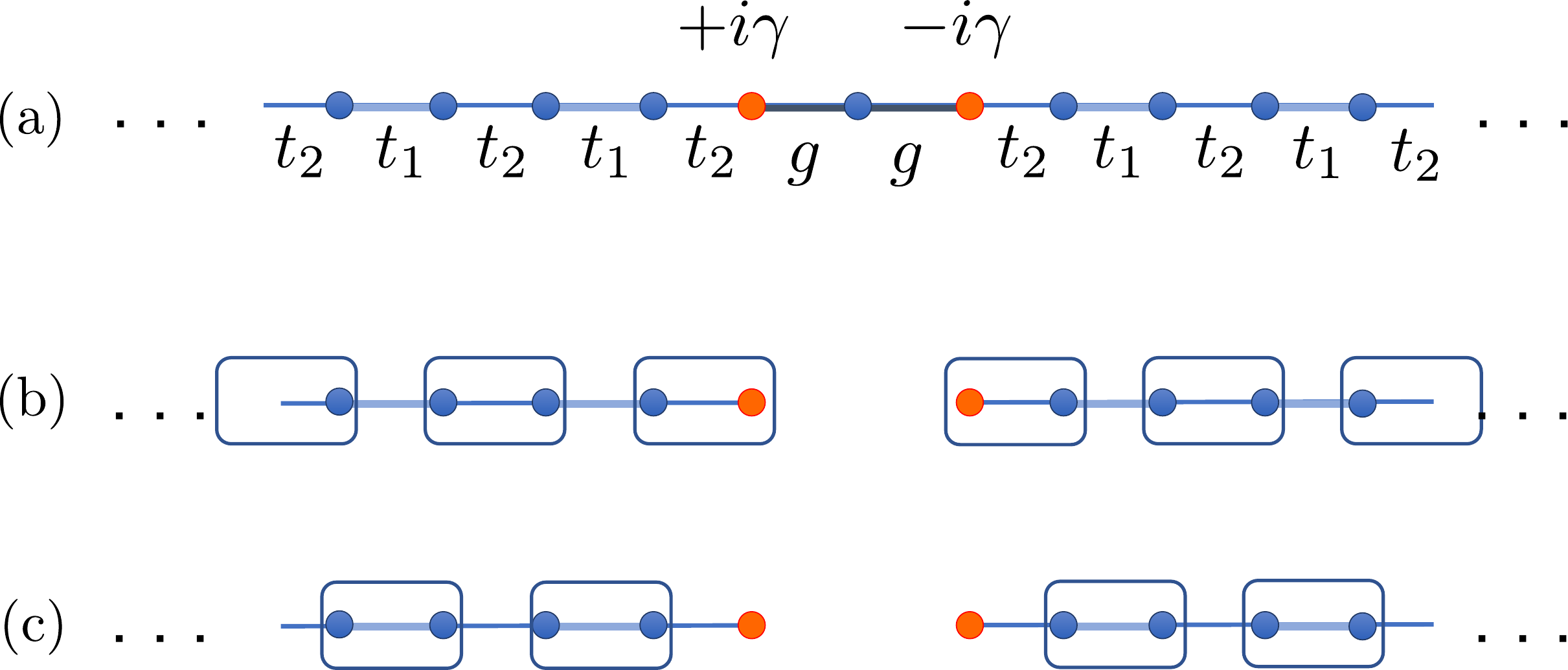}
\caption{(a) Geometry of $\PT$-symmetric open SSH model. 
The boundaries of the semi-infinite SSH chains in (b) the topologically trivial phase with $t_1<t_2$
and (c) the topologically non-trivial phase with $t_1>t_2$.}
\label{fig.1}
\end{figure*}

In Sec.~\ref{sec.eigenvalue}, we study our model under the boundary condition 
of outgoing waves to obtain the discrete energy spectra and eigenstates. This includes four discrete eigenvalues that are given as the solution to a quartic polynomial and, as shown in Sec. \ref{subsec.zero}, two states with energy eigenvalue zero that always reside exactly between the two SSH bands.
In Sec.~\ref{subsec.RAPTB}, we show the properties of the system in the reservoir-assisted broken $\PT$-symmetry regime by focusing on a parameter region in which a gap appears between this regime and the ordinary $\PT$-symmetry breaking.  This gap can be considered an example of the so-called {\it re-entrant $\PT$-symmetric phase} that has appeared in the literature \cite{JB12}.
We then analyze the situation where this gap closes in Sec. \ref{subsec.EP4} and show that it is associated with the intersection of several exceptional point surfaces.  Along the intersections, pairs of eigenvalues coalesce with one or the other zero-energy state.  In Sec. \ref{sec.surv.prob} we present simulations for the dynamics of an initially-prepared state and propose two experiments: one to verify the reservoir-assisted symmetry breaking and the other to confirm the coalesced zero-energy states.
We summarize this work and make concluding remarks in Sec.~\ref{sec.conclude}.


\section{Open system Hamiltonian and continuum dispersion}\label{sec.Hamiltonian}

The Hamiltonian of our $\PT$-symmetric open SSH model shown in Fig.~\ref{fig.1} (a)
is given by
\begin{eqnarray}
\hat{H}&=& \hat{H}_{\rm SSH} + \HPTop 
\label{ham}
\end{eqnarray}
in which
\begin{eqnarray}
\HPTop
&=&g\left(
|0\rangle \langle 1,a| + |1,a\rangle\langle 0|\right.\nonumber\\
&&\,\,\,\,\,\,\,\,\,\left. +|0\rangle \langle -1,a| + |-1,a\rangle\langle 0|
\right)\nonumber\\
&&+i\gamma \left(
|-1,a\rangle \langle -1,a| -|1,a\rangle \langle 1,a|
\right),\label{eq.H_impurity}
\label{HPT}
\end{eqnarray}
is the $\PT$-symmetric matrix Hamiltonian from Eq. (\ref{HPT.matrix}), 
written in abstract form, 
while
\begin{widetext}
\begin{eqnarray}
\hat{H}_{\rm SSH}&=& \sum_{n=1}^{\infty}
\left[
t_1\left(
|n,b\rangle \langle n+1,a| + |n+1,a\rangle\langle n,b|
\right)
+t_2\left(
|n,a\rangle \langle n,b| + |n,b\rangle \langle n,a|
\right)
\right]\nonumber\\
&&+ \sum_{n=-\infty}^{-1}
\left[
t_1\left(
|n,b\rangle \langle n-1,a| + |n-1,a\rangle\langle n,b|
\right)
+t_2\left(
|n,a\rangle \langle n,b| + |n,b\rangle \langle n,a|
\right)
\right]
\label{SSH}
\end{eqnarray}
\end{widetext}
is our reservoir Hamiltonian, consisting of two semi-infinite SSH chains.
In these equations, $|n,a\rangle$ and $|n,b\rangle$ represent sites within any part of the system
appearing at unit cell $n$ with sublattice labels $a$ and $b$, respectively.
Within $\hat{H}_{\rm SSH}$,
$t_1$ is the intercell hopping and $t_2$ is the intracell hopping.
Note that the sub-Hamiltonian $\hat{H}_{\rm SSH}$ contains the direct coupling between the SSH chains and the two $\PT$-symmetric sites.

It is well known that the SSH model can exhibit either a topologically trivial phase for $t_1 < t_2$ or non-trivial phase for $t_1 > t_2$ \cite{Asboth2016}, which can be understood for our semi-infinite chains as follows.  
In the trivial phase $t_1 < t_2$, the eigenstates are delocalized,
becoming exact dimers in the case $t_1 = 0$
as illustrated in Fig.~\ref{fig.1} (b).  
Meanwhile, a localized state appears at the end of the chain for $t_1 > t_2$ that completely decouples for the case $t_2 = 0$ as shown in Fig.~\ref{fig.1} (c).
This is the well-known edge state for the SSH model with energy $E=0$ that is interpreted as a topological invariant \cite{Asboth2016}.  
In this work, we will primarily focus our interest on the topologically non-trivial phase $t_1 > t_2$ as this is the case in which the reservoir-assisted $\PT$-symmetry breaking 
effect can be most clearly distinguished.  
We note that the authors of Ref. \cite{Kong2020} have recently studied the influence of a $\PT$-symmetric potential on the topological state in a model with a geometry somewhat similar to ours, but for which the SSH chains remain finite. 
Ref. \cite{BPCG19} meanwhile gives one of the few examples of a study in which the SSH chains are treated as open reservoirs, but for a Hermitian system in that work.
Finally, we note the photonic lattice experiment in Ref. \cite{expt1} shares some qualitative features with our geometry above (we remark on this experiment further throughout this work).

Solving the Schr\"{o}dinger equation $\hat{H}|\psi\rangle = z |\psi\rangle$ 
with eigenvalue $z$ and eigenstate $|\psi\rangle$ for Eq. (\ref{ham})
yields a series of coupled equations 
for the site amplitudes in our model.  For convenience,
we write the site amplitudes as
\begin{eqnarray}
&& \phi_0 \equiv \langle 0 |\psi\rangle, \\
&& \psi_{n,x} \equiv \langle n,x|\psi\rangle,
\end{eqnarray}
with $x = a$ or $b$.
We then multiply $\hat{H}|\psi\rangle = z |\psi\rangle$ 
on the left by $\langle 0|$ or $\langle n,x |$,
to obtain the coupled equations
\begin{eqnarray}
&& g\psi_{1,a} +  g\psi_{-1,a} = z \phi_0,\\
&& g\phi_0 + t_2 \psi_{1,b} -i\gamma \psi_{1,a} = z\psi_{1,a},\\
&& g\phi_0 + t_2 \psi_{-1,b} +i\gamma \psi_{-1,a} = z\psi_{-1,a}
\end{eqnarray}
for the amplitudes in the central $\PT$-symmetric portion of the model,
as well as
\begin{eqnarray}
&& t_2 \psi_{n,a} + t_1\psi_{n+1,a} = z\psi_{n,b}\,\,\,\,\, (\mbox{for $n\ge 1$}),\label{chain.Rnb} \\
&&t_2 \psi_{n,b} + t_1 \psi_{n-1,b} = z\psi_{n,a}\,\,\,\,\, (\mbox{for $n\ge 2$}),\label{chain.Rna}  \\
&&t_2\psi_{n,a} + t_1 \psi_{n-1,a} = z \psi_{n,b}\,\,\,\,\, (\mbox{for $n\le -1$}),\label{chain.Lnb}  \\
&&t_2\psi_{n,b} + t_1 \psi_{n+1,b} = z \psi_{n,a}\,\,\,\,\, (\mbox{for $n\le -2$}) \label{chain.Lna} 
\end{eqnarray}
for the site amplitudes within the SSH reservoirs.

The solutions to Eqs. (\ref{chain.Rnb})--(\ref{chain.Lna}) yield the two continuum eigenvalue dispersions 
\begin{eqnarray}
 z & = & \pm \sqrt{t_1^2 + t_2^2 + 2 t_1 t_2 \cos k}		\nonumber	\\
   & = &	\pm \sqrt{t_2 + t_1 e^{ik}}\sqrt{t_2 + t_1 e^{-ik}}.
\label{eq.dispersion}
\label{cont.disp}
\end{eqnarray}
These two continua or bands define the SSH reservoirs in our present model.  For the case $t_1 > t_2$, the energy for these two bands 
extend over 
the ranges from $t_1-t_2$ to $t_1 + t_2$ and from $-(t_1+t_2)$ to $-(t_1-t_2)$,
as $k$ varies along the domain $k \in \left[ 0, \pi \right]$.  
Note that an 
energy {\it band gap } 
occurs between the two bands in the range $t_1-t_2$ to $-(t_1-t_2)$, but this is entirely distinct from the gap in which the {\it discrete} eigenvalues transition from complex to real that will be introduced later in this paper.


\section{Complex eigenvalue spectrum}\label{sec.eigenvalue}

\subsection{Discrete eigenvalues under outgoing wave boundary condition}\label{subsec.eigenvalue}

To obtain the discrete eigenvalues for the Hamiltonian $\hat{H}$ given in Eq. (\ref{ham}),
we apply the Siegert boundary
condition
with outgoing waves \cite{HSNP08,Gamow28,Siegert39,GGH15,Hatano13} 
from the central $\PT$-symmetric impurity region in the form
\begin{eqnarray}
 \left(
\begin{array}{c}
\psi_{n,a}\\
\psi_{n,b}
\end{array}
\right) = e^{ikn}
\left(
\begin{array}{c}
C_a \\
C_b
\end{array}
\right)\,\,\,\,\, \mbox{for $n>0$,}
\label{outgoing.right}
\end{eqnarray}
and 
\begin{eqnarray}
 \left(
\begin{array}{c}
\psi_{n,a}\\
\psi_{n,b}
\end{array}
\right) = e^{-ikn}
\left(
\begin{array}{c}
B_a \\
B_b
\end{array}
\right)\,\,\,\,\, \mbox{for $n<0$,}
\label{outgoing.left}
\end{eqnarray}
where the wave number
$k$ is in general a complex number.
For ${\rm Im} k >0$
the wave function is localized around the central impurity region;
for ${\rm Im} k <0$ the wave function diverges into the leads for increasing $|n|$.
When the corresponding energy eigenvalue $z$ becomes complex, 
this indicates the $\PT$-symmetry of the system is broken.
Applying the boundary conditions (\ref{outgoing.right}) and (\ref{outgoing.left})
to the coupled equations previously derived in Sec. \ref{sec.Hamiltonian} 
we obtain equations for the amplitudes $C_a,C_b,B_a,B_b$ and $\phi_0$ in the impurity region as
\begin{eqnarray}
 && ge^{ik}(C_a + B_a) = z\phi_0,\\
 && g\phi_0 e^{-ik} + t_2 C_b = (z+i\gamma) C_a,\\
 && g\phi_0 e^{-ik} + t_2 B_b = (z-i\gamma) B_a,
\end{eqnarray}
and within the SSH leads as
\begin{eqnarray}
&& (t_2 + t_1 e^{ik}) C_a =z C_b,  \label{chain.Rnb.C} \\
&& (t_2 + t_1 e^{-ik}) C_b = z C_a, \label{chain.Rna.C}\\
&& (t_2 + t_1 e^{ik}) B_a = z B_b, \label{chain.Lnb.C}\\
&& (t_2 + t_1 e^{-ik}) B_b = z B_a. \label{chain.Lna.C}
\end{eqnarray}
The latter equations yield the SSH continuum eigenvalues previously reported in Eq. (\ref{cont.disp}).
Meanwhile we can eliminate $C_b$ and $B_b$ from the first three equations to obtain 
\begin{widetext}
\begin{eqnarray}
 \left(
\begin{array}{ccc}
 \pm t_2 \frac{\sqrt{t_2+t_1e^{ik}}}{\sqrt{t_2+t_1
  e^{-ik}}}-i\gamma-z&g e^{-ik} &  0\\
ge^{ik} & -z & ge^{ik}\\
0&ge^{-ik} &  \pm t_2 \frac{\sqrt{t_2+t_1e^{ik}}}{\sqrt{t_2+t_1
  e^{-ik}}}+i\gamma-z
\end{array}
\right)
\left(
\begin{array}{c}
 C_a\\
\phi_0\\
B_a
\end{array}
\right)=0.
\label{Heff.3x3}
\end{eqnarray}
The determinant of the matrix in the left hand side is given by
\begin{eqnarray}
D_s(\lambda) \equiv \mp \frac{\sqrt{t_2 + t_1\lambda}}{\sqrt{t_2 + t_1/\lambda}}
\left\{
t_1^2 (t_2+t_1\lambda)/\lambda^{2}+\gamma^2(t_2 + t_1/\lambda) -2 g^2 t_1/\lambda
\right\},
\end{eqnarray}
in which we have defined
\begin{eqnarray}
\lambda = e^{ik}. 
\label{lambda}
\end{eqnarray}
The solution of the equation $D_s(\lambda_j)=0$ can be obtained as
\begin{eqnarray}
 \lambda_{\pm} = \frac{1}{2\gamma^2 t_2}\left\{
-t_1(t_1^2+\gamma^2-2g^2)\pm \sqrt{t_1^2(t_1^2 + \gamma^2 -2g^2)^2-4t_1^2t_2^2\gamma^2}
\right\},\label{eq.lambda_pm}
\end{eqnarray}
\end{widetext}
which then yields the wave number $k$ of the discrete eigenvalues as
\begin{eqnarray}
 k = -i \log \lambda_\pm  \label{eq.k_lambda}
\end{eqnarray}
after inverting Eq. (\ref{lambda}).
Using Eq. (\ref{cont.disp}) the energies of the discrete eigenvalues 
can be obtained in terms of $\lambda_{\pm}$ as
\begin{eqnarray}
 z = \pm \sqrt{t_2 + t_1 \lambda_\pm}\sqrt{t_2+t_1/\lambda_\pm}.
\end{eqnarray}
Squaring this equation twice and applying Eq. (\ref{eq.lambda_pm}) 
we can obtain the polynomial equation for the energy eigenvalues $z_j$ directly in terms of the system parameters as
\begin{eqnarray}
 P_s(z_j) = 0,
\end{eqnarray}
where $P_s(z)$ is the biquadratic polynomial
\begin{eqnarray}
P_s(z) &\equiv&
 \gamma^2 z^4 + \left\{
\gamma^4 -2\gamma^2(t_2^2+g^2) + t_1^4 -2g^2t_1^2
\right\}z^2\nonumber\\
&&+(t_1^2 -t_2^2-2g^2)
\left\{
\gamma^2(t_1^2-t_2^2)-2g^2t_1^2
\right\}.
\label{eq.P_s(z)}
\label{poly}
\end{eqnarray}
We emphasize that the solutions of $P_s(z_j)=0$ are equivalent to the previous equation $D_s(\lambda_j) =0$.
Further, these four solutions always appear as two pairs for which the members of each pair satisfy $z_j = - z_l$, which is inherited from the chiral symmetry properties of the underlying SSH model.


\subsection{Zero-energy modes}\label{subsec.zero}

\begin{figure}
\includegraphics[width=0.92\columnwidth]{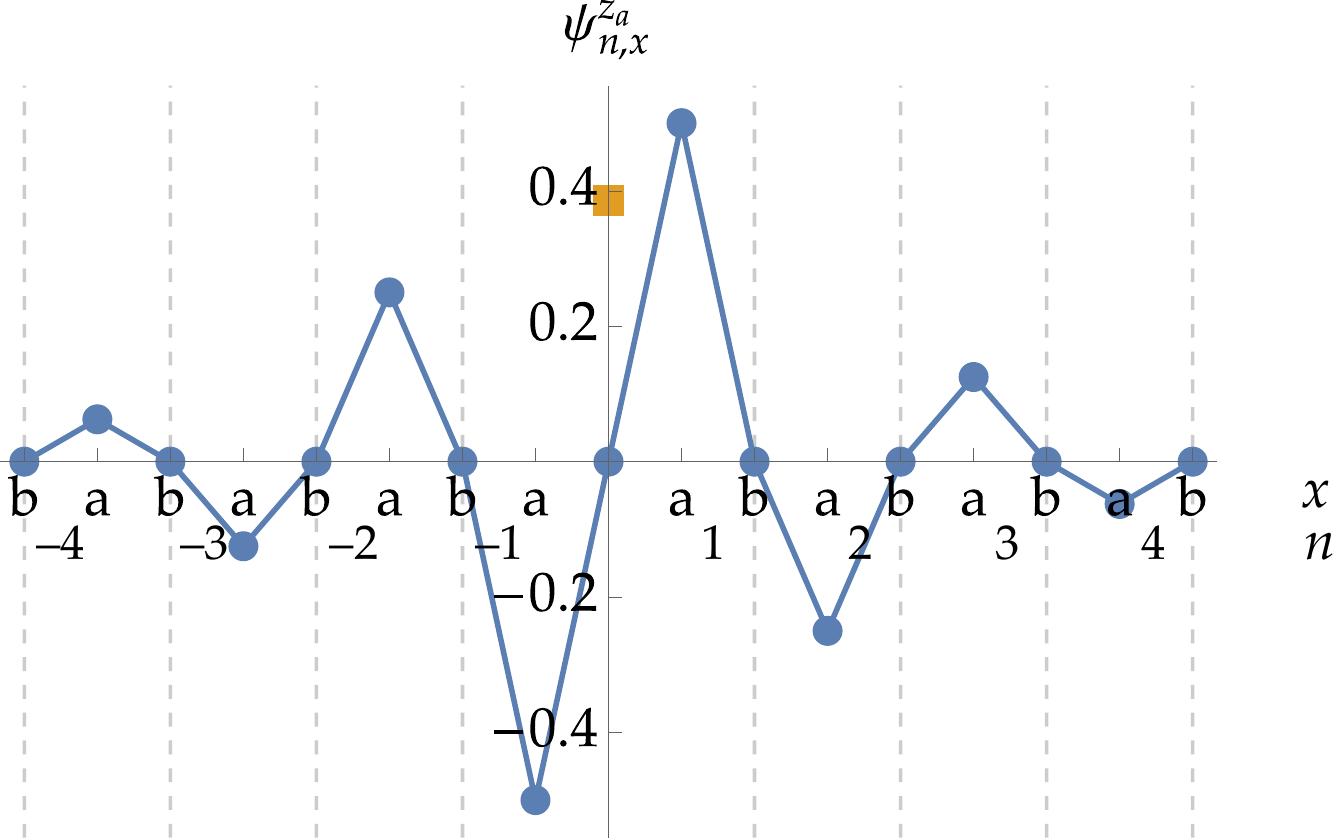} 
\\
\hspace*{0.04\columnwidth}(a)\hspace*{0.96\columnwidth}
\\
\includegraphics[width=0.93\columnwidth]{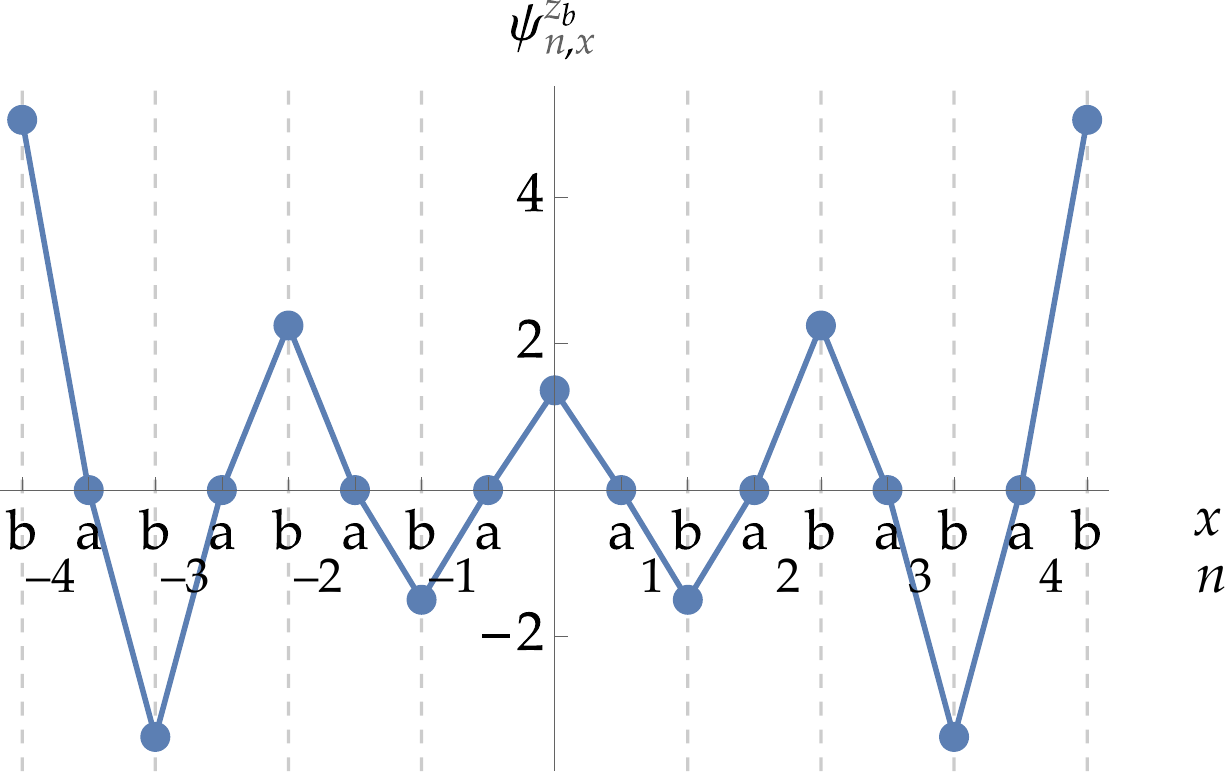} 
\\
\hspace*{0.04\columnwidth}(b)\hspace*{0.96\columnwidth}
\caption{ 
Wave function amplitudes of the zero-energy eigenstates for the case $t_1 > t_2$. 
(a) Localized state $|\psi^{z_a}\ket$ with $t_1 = 2$ and $C_a = -1$; and (b) anti-localized state $|  \psi^{z_b} \ket$ with $t_1 = 1.5$ and $C_b = 1$.  The other parameters used are $t_2 = 1$, $g =1.1$ and $\gamma = 0.85$ in both panels.  The alternating $b$ sites are emphasized with dashed vertical lines.
Note the lone non-zero imaginary part of the wave function for $\phi^{z_a}_{0}$ is indicated with an orange square in (a), corresponding to Eq. (\ref{za.0}).
}
\label{fig:zero}
\end{figure}

It can be shown there are two further discrete solutions not yet accounted for in the previous analysis, both of which have eigenvalue $z=0$ (residing directly in between the two SSH bands).
Setting $z=0$ in Eqs. (\ref{chain.Rnb.C})--(\ref{chain.Lna.C}), we find there are two non-trivial cases.
For one of the solutions the condition 
\beq
\lambda = e^{ik_a} = - t_2/t_1
\label{ka.cond}
\eeq
must be satisfied, which gives 
$k_a = \pi + i \log (t_1/t_2)$
and further requires
\beq
  C_a = - B_a \neq 0, \ \ \ \ \ \ \ \ \  C_b = B_b = 0
\label{za.C}
\eeq
as well as
\beq
  \phi_0 = - i \frac{\gamma t_2}{g t_1} C_a
\label{za.0}
\eeq
for the central site $| 0 \ket$.  Since this solution is non-zero on the $a$ sites, we refer to it as 
\beq
|  \psi^{z_a} \ket
\label{psi.za}
\eeq
with eigenvalue $z_a = 0$.
The condition for the other solution is 
\beq
\lambda = e^{ik_b} = - t_1/t_2
	,
\label{kb.cond}
\eeq
yielding $k_b = \pi + i \log (t_2/t_1)$ and
\beq
  C_a = B_a = 0, \ \ \ \ \ \ \ \ \  C_b = B_b \neq 0
\label{zb.C}
\eeq
as well as
\beq
  \phi_0 = \frac{t_1}{g} C_a
  	.
\label{zb.0}
\eeq
We refer to this solution as 
\beq
|  \psi^{z_b} \ket
\label{psi.zb}
\eeq
with eigenvalue $z_b = 0$.
In the case $t_1 > t_2$ that is the primary focus of this paper, 
the mode $|\psi^{z_a}\ket$ 
represents a localized state with odd wave function in the SSH leads of 
the form $\psi^{z_a}_{\pm n,a} = \bra \pm n,a|\psi^{z_a}\ket = \pm (-t_2/t_1)^{|n|} C_a$ 
(with $\psi^{z_a}_{n,b} =\bra n,b|\psi^{z_a}\ket= 0$ for all $n$), as shown in Fig. \ref{fig:zero}(a).
Meanwhile, $z_b$ is instead an anti-localized state with even wave function in the leads of the form $\psi^{z_b}_{\pm n,b} = (-t_1/t_2)^{|n|} C_b$ (with $\psi^{z_b}_{n,a} = 0$ for all $n$), as shown in Fig. \ref{fig:zero}(b).  For $t_2 > t_1$, the two states switch roles in terms of localization properties.

The form of these two wave functions appearing in terms of powers of $-t_1/t_2$ with alternating vanishing amplitudes recalls the same properties of the edge states in the finite SSH model \cite{Asboth2016,PLA1}. 
Further, the localized solution
$|\psi^{z_a}\ket$ 
seems to be roughly comparable to the $\PT$-symmetric topological interface state observed in the experiment in Ref. \cite{expt1}.
However, to be careful, we refer to 
$|\psi^{z_a}\ket$ and $|\psi^{z_b}\ket$ 
as {\it zero-energy states} for the remainder of this work.

For now we turn to the evaluation of the remaining four discrete eigenstates coming from the solutions of the polynomial equation $P_s (z) = 0$.  However, we note there are some special circumstances in which pairs of solutions from these four eigenstates can also become zero-energy eigenstates, in which case those pairs will coalesce with one or the other of the pre-existing zero-energy states.  For this reason, we never explicitly plot the above two zero-energy eigenvalues in the figures that follow (because that would unfortunately obscure the special cases in which the other solutions also become zero-energy states).
We further insert a slight note of caution on this point that, technically, the boundary conditions in Eqs. (\ref{za.C}) and (\ref{zb.C}) are different than the boundary conditions originally used to derive the four polynomial solutions starting from Eqs. (\ref{outgoing.right}) and (\ref{outgoing.left}).  However, given the self-consistency of all the results obtained in this work, it seems safe to treat the coalesced zero-energy states as a limiting case of the other solutions.  We further present a quick self-consistency check on the {\it physical} results later in Sec. \ref{sec.surv.prob}.


\subsection{Reservoir-assisted $\PT$-symmetry breaking}\label{subsec.RAPTB}

\begin{figure*}[t]
  \includegraphics[width=18cm]{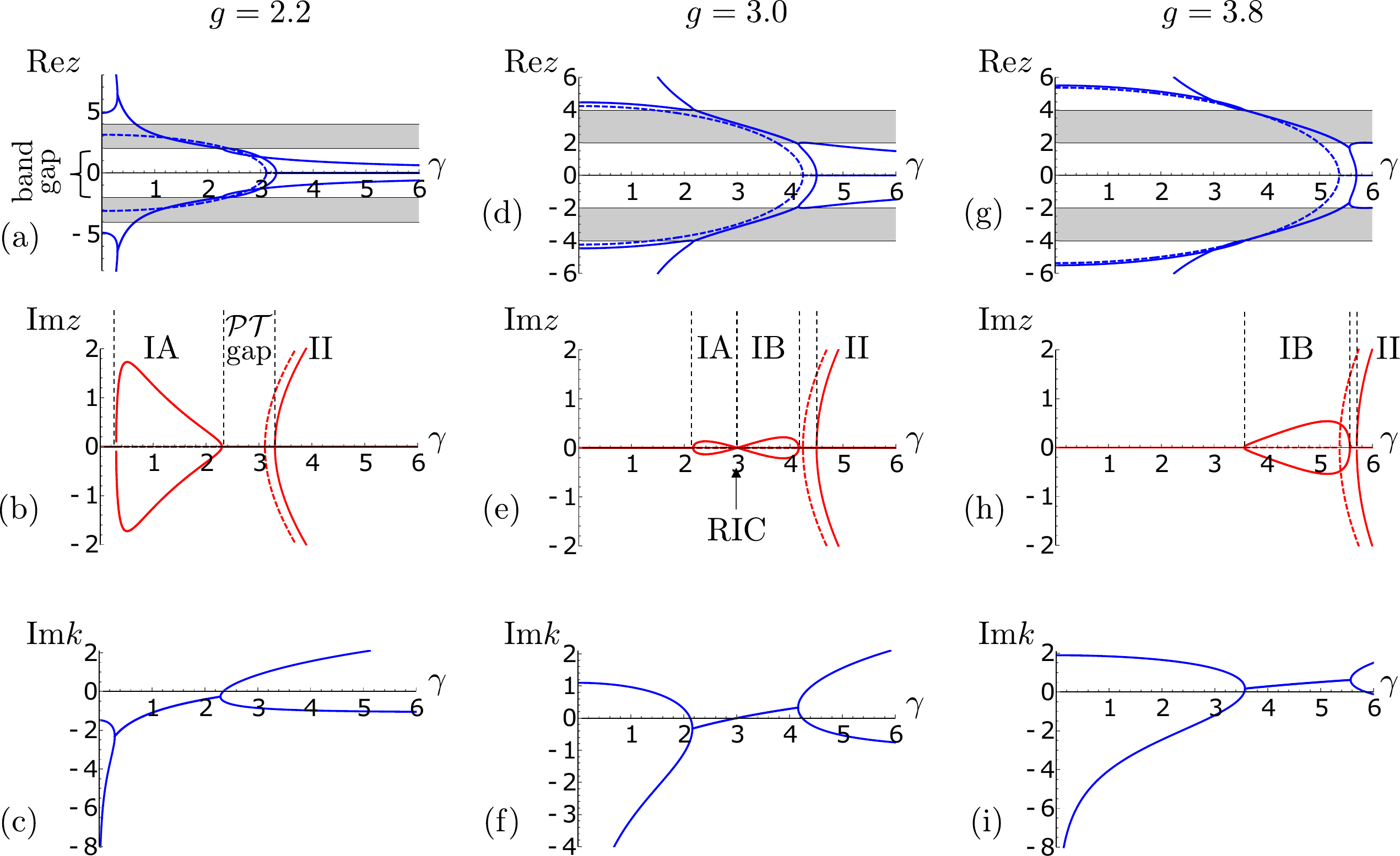}
 \caption{
Discrete eigenvalue spectrum of the $\PT$-symmetric open SSH model 
for $t_1 = 3.0$
in the unit of $t_2 = 1$ (solid lines):
(a) real and (b) imaginary parts of energy and (c) imaginary part of wave number 
for $g=2.2$,
(e) real and (f) imaginary parts of energy and (g) imaginary part of wave number 
for $g=3.0$,
and
(h) real and  (i) imaginary parts of energy and (j) imaginary part of wave number 
for $g=3.8$.
In (a), (d), and (g), the shaded regions are the energy continua of the SSH model.
For comparison, the two complex eigenvalues from the decoupled Hamiltonian $H_{\PT}$ are also shown by broken lines in the first two rows.
The resonance in
 the continuum(RIC) appears 
at $\gamma=3.0$ 
(where ${\rm Im} E =0$ and ${\rm Im} k=0$) in the $g=3.0$ case.
Note that we evaluate the wave number in the unit where the lattice constant is unity.
}\label{fig.2}
\end{figure*}

We now turn to the four remaining discrete eigenvalues from the quartic polynomial equation $P_s (z) = 0$.
To illustrate the most interesting cases, we plot these eigenvalues
for $g=2.2$ (left column), 
$g=3.0$ (middle column) and $g=3.8$ (right column) in Fig.~\ref{fig.2}.
We present the real part of the discrete energy eigenvalues $z_j$ in the upper row
 and the imaginary part of $z_j$ in the middle row of this figure  [full lines].
The imaginary parts of the associated wave number $k_j$ are also shown 
in the lower row of Fig.~\ref{fig.2}.
The first observation we make is that, 
for each of these cases in Fig.~\ref{fig.2} (b), (e) and (h),
as the strength of the complex potential $\gamma$ increases
the imaginary part of the energy eigenvalue first appears at a non-zero value of $\gamma$, then vanishes again as we further increase $\gamma$,
and then once again reappears.
We divide the parameter range in which complex eigenvalues exist into two regions:
the lower region with relatively smaller values of $\gamma$ is Region I; while the second, appearing for larger values of $\gamma$, is Region II.
Notice that Region I occupies a finite range of the $\gamma$ parameter space with both a lower and upper boundary, while Region II is semi-infinite, having only a lower boundary.
An EP marks each of these three boundaries.
Note that later we will further subdivide Region I into Regions IA and IB according to the respective eigenstate properties. 

An immediate and non-trivial observation from Fig.~\ref{fig.2} is that the $\PT$-symmetry, once broken at the lower threshold of Region I, is recovered again at the upper threshold of Region I as the strength of the non-Hermitian parameter $\gamma$ is increased.  
This counterintuitive result contrasts with the usual picture in isolated
$\PT$-symmetric systems for which, once broken, the $\PT$-symmetry is usually never restored.
We make this comparison explicit by plotting the eigenvalues $z_\pm = \pm \sqrt{2g^2 - \gamma^2}$ from the isolated $\PT$-symmetric Hamiltonian $\HPT$ [Eq. (\ref{HPT.matrix})] as the broken lines in the upper two rows of Fig.~\ref{fig.2}.  
For the case $g=2.2$ 
illustrated in Fig. \ref{fig.2}(b) 
we see that the entire domain of Region I falls well below the $\PT$-symmetry breaking threshold for the isolated system $\HPT$,
which occurs at $\gPT = \sqrt{2}g \approx 3.111$.
Meanwhile the exceptional point threshold for the complex eigenvalues in Region II is shifted a bit above that of the isolated system $\HPT$.  Hence, compared to the isolated $\PT$ system,
the presence of the reservoir in this case provides both a destabilizing effect (due to the appearance of Region I) and a stabilizing effect (by shifting upward the threshold for Region II).

We also emphasize that the gap between Regions I and II in which the reality of the eigenvalues is restored [see Fig.~\ref{fig.2}(b)] is entirely different from the band gap between the two SSH energy bands [see Fig.~\ref{fig.2}(a)].  To distinguish these, we will refer to the former as a ``$\PT$ gap'' or simply a ``gap'' throughout the paper, while we will always refer to the latter as a ``band gap.''

We next make the comparison between Region II and the isolated $\PT$-symmetric system more explicit in the following two points.  First, we can find the exact value of the exceptional point marking the lower boundary of Region II by seeking values of the parameter $\gamma$ that yield a double root of the polynomial equation (\ref{poly}).  This can be obtained from the solutions of the discriminant equation $D_P = 0$ in which the discriminant of Eq. (\ref{poly}) is given by (ignoring an uninteresting overall factor)
\beqa
  D_P & = & (t_1^2 - \gamma^2)^4 \left( 2g^2 t_1^2 + (t_2^2 - t_1^2) \gamma^2 \right)	\nonumber  \\
  	& &\times\left[ (t_1^2 - 2g^2)^2 - 2 (2g^2 - t_1^2 + 2t_2^2) \gamma^2 + \gamma^4 \right]^2. \ \  
\label{poly.disc}
\eeqa
The second factor in this expression yields the relevant exceptional point that marks the boundary for Region II at $\gamma = \gRII$, in which
\beq
  \gRII 
  	= \gPT \frac{t_1}{\sqrt{t_1^2 - t_2^2}}
	,
\label{EPII}
\eeq
written in terms of the original EP from the isolated $\PT$ system $\gPT = \sqrt{2} g$.  
Second, we comment on the eigenstates in Region II.  In the bottom row of Fig. \ref{fig.2}, we see that at least two of the Region II eigenstates always have a positive imaginary part of the $k$ value, which means these states are localized in the region of the central $\PT$-symmetric impurity.  More specifically, as $\gamma$ becomes quite large $\gamma \gg \gRII$, the wave function for these two eigenstates increasingly becomes localized on either of the $| \mp 1, a \ket$ sites, while their eigenvalues become approximately $\approx \pm i \gamma$, respectively.  Hence, for large $\gamma$, these two states increasingly act like the bare non-Hermitian impurities from the isolated $\HPT$ Hamiltonian.

Before turning to the properties of the eigenstates in Region I, we briefly comment in more detail on the characteristics of the EP at $\gamma = \gRII$.  In the immediate vicinity of the EP, we can expand the eigenvalues in the characteristic Puiseux expansion \cite{Kato,GraefeEP3,GRHS12} that can be obtained as
\beqa
  z_{\textrm{II},\pm} 
   &	= & \pm \frac{1}{t_1} \sqrt{ \frac{(t_1^2 - t_2^2)^3}{2 g^2 t_2^2 - t_1^2 (t_1^2 - t_2^2)} }
			\left( \gamma^2 - \gRII^2 \right)^{1/2}
				\nonumber		\\
   & &		+ \mathcal{O} \left( \gamma^2 - \gRII^2 \right)
  	.
\label{EPII.puiseux}
\eeqa
We note that this form is typical of a second-order exceptional point (EP2), involving two coalescing eigenstates.  However, notice that exactly at the EP, these two eigenvalues coincide at $z=0$.  
We further find that the value of $k$ exactly at the EP is determined by the condition $\lambda = e^{ik} = - t_2/t_1$, which coincides with the condition (\ref{ka.cond}) that yielded the localized zero-energy mode $z_a = 0$ with the wave function coefficients determined from Eq. (\ref{za.C}).  This suggests that the two eigenstates with eigenvalues $z_{\textrm{II},\pm}$ are actually coalescing with this pre-existing zero-energy state at $\gamma = \gRII$, giving rise to a third-order exceptional point (EP3).  We note the possibility of just such an apparent mismatch between the order of the EP and its Puiseux expansion is discussed in detail in Ref. \cite{GraefeEP3} (further, see the chiral symmetry-protected EP3 in Ref. \cite{Mandal21}, which is also a zero-energy mode that behaves similarly).  We further note that the time evolution simulation discussed in Sec.~\ref{sec.surv.prob}
strongly suggests that the order of the EP is indeed three.

\begin{figure*}[t]
  \includegraphics[width=18cm]{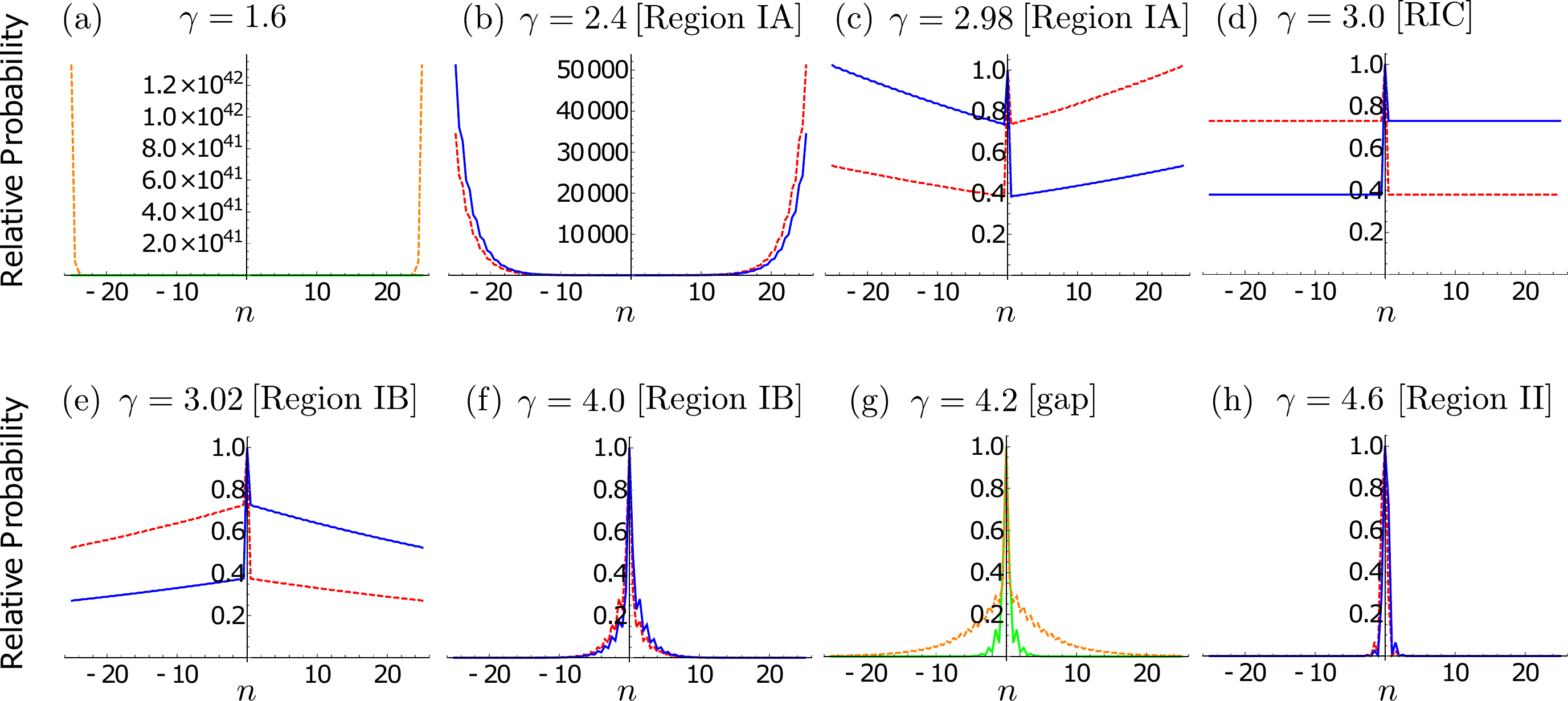}
 \caption{
Square of the absolute value of eigenfunctions
 at $n$th site $|\psi_{n,x}|^2$ 
when the eigenfunctions are 
normalized as $|\phi_0|^2 =1$ (relative probabilities)
for 
(a) $\gamma=1.6$, 
(b) $\gamma=2.4$,
(c) $\gamma=2.98$,
(d) $\gamma=3.0$,
(e) $\gamma=3.02$,
(f) $\gamma = 4.0$,
(g) $\gamma=4.2$, and
(h) $\gamma=4.6$
in the system with
$t_1=3.0$ and $g=3.0$ in the unit of $t_2=1$
shown in Fig.~\ref{fig.2} (d), (e) and (f).
In (b), (c), (d), (e), (f), (h), 
blue solid lines represent
asymmetric eigenfunctions associated with complex energy eigenvalues with 
negative imaginary parts,
while red broken lines represent
asymmetric eigenfunctions associated with complex energy eigenvalues with 
positive imaginary parts.
In (a) and (g), orange broken lines and green solid lines represent
symmetric eigenfunctions with real energy eigenvalues.
We show relative probabilities for only two out of four eigenfunctions for each $\gamma$.
In the figures from (a) to (f), we plot probabilities only for
eigenfunctions with positive real parts of complex energy
eigenvalues because the other eigenfunctions with negative real parts 
of the energies
give the 
same probabilities.
In the figure (h),
we only show relative probabilities for the two localized eigenfunctions,
into which the two eigenfunctions corresponding to the green solid line in (g) are changed as $\gamma$ is increased,
and eliminate diverging eigenfunctions 
with real energy eigenvalues
and negative imaginary parts of the wave numbers(virtual bound states),
into which the two eigenfunctions corresponding to the orange broken line in (g) are changed as $\gamma$ is increased.
Note that we evaluate the wave number in the unit where the lattice constant is unity.}
\label{fig.3}
\end{figure*}

In contrast to Region II,
the behavior of the complex eigenvalues in Region I 
is more subtle.
Notice from the top row of Fig. \ref{fig.2} that 
the real parts of the discrete energy eigenvalues in Region I overlap
with the two energy continua of the reservoir.
This is similar to the resonance condition in traditional Hermitian open quantum systems, in which the eigenvalue becomes complex
when the energy of a quantum level resides within (or resonates with) the energy continuum
(for examples of this in Hermitian tight-binding systems, see Refs. 
\cite{HSNP08,GNHP09,HO14,DBP08,OH17A}).
For the present model, when the resonance condition is satisfied the coupling with the reservoir 
is enhanced, which results in broken $\PT$-symmetry with complex eigenvalues in Region I.
Hence, we refer to the symmetry breaking in Region I as
 {\it reservoir-assisted $\PT$-symmetry breaking}.
 However, the eigenstate properties are more complex in Region I than the traditional Hermitian case.
In Hermitian open systems, the resonant state with complex energy eigenvalue
is always associated with an  eigenfunction that is spatially divergent and hence
cannot be normalized in the usual manner.\cite{PPT91,HSNP08,Madrid12} 
By contrast, in non-Hermitian systems
eigenstates with complex eigenvalues can have either
diverging or localized wave functions.\cite{GGH15}
For the case $g=2.2$,
the eigenfunction in Region I diverges
because ${\rm Im} k<0$
as  shown in Fig.\ref{fig.2} (c).
For the case $g=3.8$, 
the eigenfunction in Region I is instead localized
because ${\rm Im} k>0$ as
shown in Fig.~\ref{fig.2} (i).
To distinguish these behaviors, we refer to the Region I parameter domain in which the eigenstates are divergent as Region IA, while referring to that in which these eigenstates are localized as Region IB.
The most interesting case is illustrated in Fig.~\ref{fig.2} (e) 
for $g=3.0$ 
in which both Regions IA and IB are present.
As seen in Fig.~\ref{fig.2} (d) (e), at the crossing point between IA and IB the eigenvalue width (imaginary part of the eigenvalue) vanishes while the real part of the eigenvalue remains within the continuum.  Hence the eigenvalue at this crossing point becomes embedded within the continuum itself.  Further, in Fig.~\ref{fig.2} (f) we see that the imaginary part of the wave vector $k$ also vanishes at this point, so that $k$ becomes purely real-valued.
We refer to this state as resonance in continuum (RIC) \cite{GGH15,Shobe2021}, as it behaves precisely like a resonance that has become stuck inside the continuum.  
This is an example of what is also known in the literature as a spectral singularity,\cite{Mostafa2009lett,Mostafa2009A,Mostafa2011,Longhi2009,Longhi10,ZK20} in which 
the eigenstate corresponding to each of these eigenvalues
is a steady state delocalized over the surrounding reservoir (we will show this more explicitly momentarily).  

The exact location of the RIC can be obtained from the first factor of the discriminant reported in Eq. (\ref{poly.disc}) and is given simply by $\gamma_\textrm{RIC} = t_1$, which is indeed consistent with  Fig.~\ref{fig.2} (e) (f).  However, we emphasize that the RIC only exists in the case that both Regions IA and IB are present in the spectrum. 
Starting from this observation, we can more precisely state the condition for the existence of the RIC as follows.  We reiterate that in Fig.~\ref{fig.2} (f) the RIC occurs when the imaginary part of the resonance wave vector $k$ vanishes as $k$ crosses the real axis.  Such a crossing can occur only if the EP marking the lower edge of Region I has imaginary part of $k$ with opposite sign to that of the EP marking the upper edge of Region I (compare Figs. ~\ref{fig.2}(c) and (i), in which the RIC does not appear, with Fig. ~\ref{fig.2}(f) in which the RIC is present).  The exact locations of these EPs can be easily obtained from the third factor of the discriminant in Eq. (\ref{poly.disc}), which gives
\beq
  \gRImp
  	= \mp t_2 + \sqrt{2g^2 + t_2^2 -t_1^2}
	.
\label{EPI}
\eeq
Here, $\gRIm$ ($\gRIp$) marks the lower (upper) edge of Region I.  Hence, when the RIC exists, it must reside in the parameter range $\gRIm < \gamma_\textrm{RIC} < \gRIp$, between the two EPs.  Further, the condition $\gRIm = \gamma_\textrm{RIC} = t_1$ ($\gRIp = \gamma_\textrm{RIC} = t_1$) denotes the point at which the RIC appears or vanishes at the lower (upper) edge of Region I.  These conditions can be solved to obtain the range of $g$ values that permit the RIC to be realized in the spectrum as
\beq
  \sqrt{t_1 (t_1 - t_2)} < g < \sqrt{t_1 (t_1 + t_2)}
  	.
\eeq

Next we turn to the gap between Regions I and II in which the $\PT$-symmetry
is restored \cite{JB12}.  We note this effect can only occur in the case $t_1 > t_2$ corresponding to the topologically non-trivial phase of the bare SSH chains 
(although not even then for some parameter values, as detailed in Sec. \ref{subsec.EP4}).
Recall that in this case ($t_1 > t_2$), the bare semi-infinite SSH leads  [Eq. (\ref{SSH})]
give rise to edge states that are localized around $| \pm 1, a \ket$.  These sites incorporate the $\PT$-symmetric potentials $\pm i \gamma$ in the coupled model Eq. (\ref{ham}), and for increasing values of $\gamma$ they have a tendency to couple much more strongly within the central $\PT$ system rather than with their respective SSH chains. 
Hence, in the parameter regime of the gap, the model tends to act more like the $\PT$-symmetric phase of the decoupled $\PT$-symmetric Hamiltonian in Eq. (\ref{HPT.matrix}).
On the other hand,
in the topologically trivial phase $t_1 < t_2$ 
this restoration of the $\PT$-symmetry never occurs, 
because the individual impurity sites couple strongly to their respective SSH chains.
Hence, in this parameter region, the system acts less like a $\PT$-symmetric system coupled with a reservoir, and more like two independent SSH chains, each with an attached non-Hermitian 
impurity.  In other words, the system acts more like a generic non-Hermitian model, for which the eigenvalues are generally complex.

Having established the spectral properties of the model in the gapped case illustrated in Fig. \ref{fig.2}, we now take a closer look at the corresponding wave function properties in Fig.~\ref{fig.3}. 
Here we plot the relative probability of the eigenfunctions at each site 
for the case $g=3.0$ [corresponding to Fig.~\ref{fig.2} (d), (e) and (f)], since every possible region can be represented in this case.
For the panels of Fig.~\ref{fig.3} corresponding to the two broken $\PT$-symmetry regions, 
the eigenfunctions corresponding to complex eigenvalues are spatially asymmetric (although anti-symmetric when taken as a pair, as demanded by $\PT$-symmetry).
This can be seen for the eigenfunctions in Region I shown in
Fig.~\ref{fig.3} (b-f) and the eigenfunctions in Region II shown in
Fig.~\ref{fig.3} (h).
The eigenfunctions diverge in Region IA 
(with $\gamma<3.0$)
as shown in Fig.~\ref{fig.3} (b) and (c),
while they are shallowly localized in Region IB
($\gamma >3.0$)
as shown in Fig.~\ref{fig.3} (e) and (f).
Meanwhile, the delocalized wave function for the RIC 
at $\gamma=3.0$
is uniform along each individual chain, while asymmetric about the system as a whole as shown in Fig.~\ref{fig.3} (d).
This picture illustrates that the RIC is a singular steady state, for which particles injected through or absorbed 
into the two non-Hermitian impurities are exactly balanced with the drain into or injection from the two leads.
Hence the RIC is an example of a non-equilibrium steady state that constantly reprocesses particles as they are transported from the central impurity region to the reservoirs, or vice versa.  
Since the eigenfunction of the RIC in Fig.~\ref{fig.3} (d) is asymmetric,
the $\PT$-symmetry of the system is broken at this point, even though 
the imaginary parts of the two eigenvalues vanish \cite{GGH15}.

Finally, as illustrated in  Fig.~\ref{fig.3} (h), as we go deeper into Region II for increasing $\gamma$, the $\PT$-symmetric wave functions become more strongly localized on the two $\PT$-symmetric impurity sites $| \pm 1, a \ket$.  Hence, for $\gamma \gg \gPT$ these wave functions increasingly mimic the properties of the original Hamiltonian $\HPT$ absent the coupling to the SSH chains.

While we expect that the reservoir-assisted $\PT$-symmetry breaking (Region I) can appear quite generally
in $\PT$-symmetric open quantum systems,
in most cases it would probably be more difficult to cleanly distinguish
this from the ordinary $\PT$-symmetry breaking (Region II).\cite{GGH15}
There are two reasons for this:
one is that 
the two mechanisms for the appearance of complex eigenvalues are 
usually mixed;
the other is that there is not a qualitative difference between the eigenfunctions in Region IB and Region II.
A key advantage of our open SSH model is that the topological properties of the SSH chains helped in creating a gap between Region I and II in much of the parameter space.

We briefly note that the resonance and anti-resonance with complex conjugate eigenvalues and localized wave functions in Regions IB and II are a peculiar feature that is particular to $\PT$-symmetric open quantum systems.  We comment on this point further in Sec.~\ref{sec.conclude}.


\section{Gap closings and coalesced zero-energy modes}\label{subsec.EP4}

\begin{figure*}[t]
  \includegraphics[width=10cm]{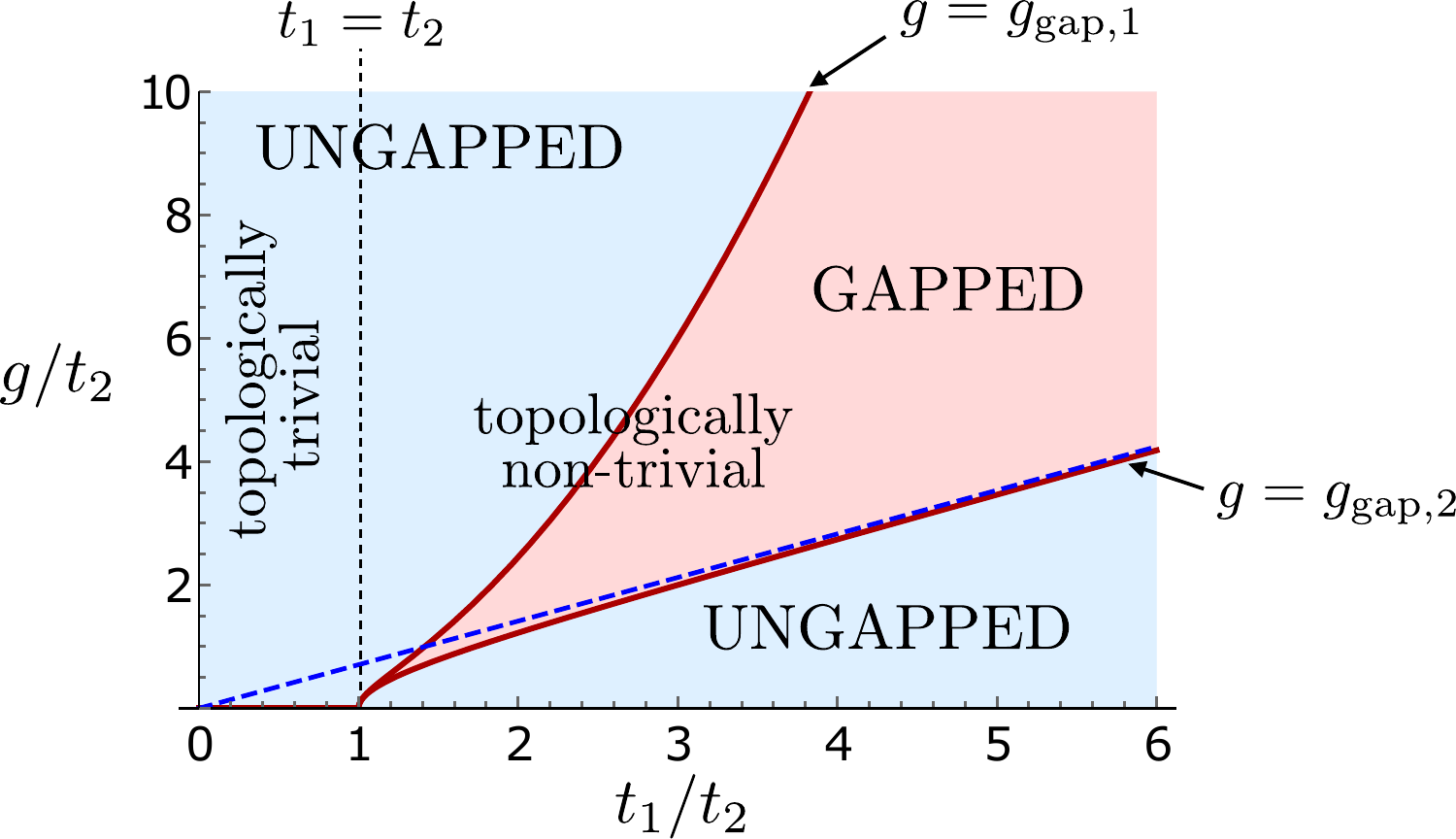}
 \caption{Phase diagram for the gap closing between Region I and II in the $t_1/t_2$-$g/t_2$ plane. 
The two dark red solid lines represent
the phase boundaries between the gapped and the ungapped regions
obtained from Eq.~(\ref{EP4.cond}).  
The dashed blue line represents the line $g = t_1/\sqrt{2}$ along which $\gRIm$ vanishes.  For $g >  t_1/\sqrt{2}$, the condition $\gRIm = - t_2 + \sqrt{2g^2 + t_2^2 -t_1^2}$ gives the lower boundary of Region I.  But for $g <  t_1/\sqrt{2}$, we instead have $\gRIm = \left| - t_2 + \sqrt{2g^2 + t_2^2 -t_1^2} \right| $ that represents an EP2 for which the $\PT$-symmetry is broken on either side of the exceptional point (in other words, below the blue dashed line, the $\PT$ symmetry is broken for an infinitesimally small value of $\gamma$).
}
\label{fig.phase}
\end{figure*}

In the preceding analysis, we stated that the $\PT$ gap between Region I and Region II only exists in the case corresponding to the topologically non-trivial phase of the decoupled SSH chains, and even then only under certain conditions.  Here we make this statement more precise by showing that the gap closes for either sufficiently large or small values of $g$ in terms of the other parameters, which gives rise to both an upper and a lower boundary in the parameter space of the model [see the red lines in Fig. \ref{fig.phase}].  We further show that these gap closings occur along the intersection of several exceptional surfaces and again involve the zero-energy modes.
Curiously, the qualitative features of the two boundaries are rather different.

\begin{figure}
\includegraphics[width=0.92\columnwidth]{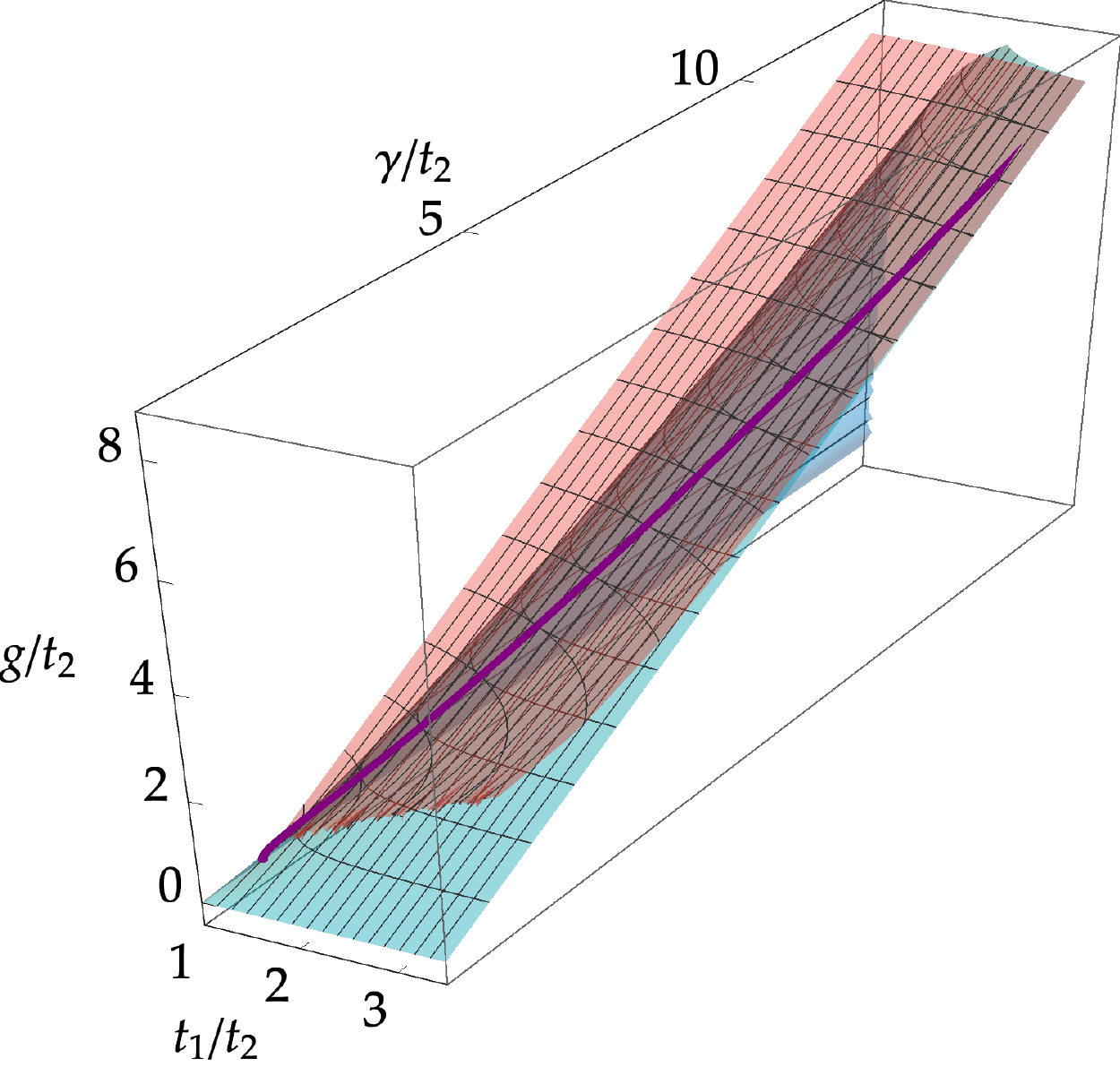} 
\caption{EP2 surfaces $\gamma = \gRIp$ (red) and $\gamma = \gRII$ (blue) are shown 
in the $(t_1/t_2, g/t_2, \gamma/t_2)$-parametric space of the model.
The EP5 line along $g = \ggap$ and $\gamma = \gEPfp = t_1^2/t_2$ is shown as the purple curve occurring along the intersection of the two EP2 surfaces \cite{ENexus}.  The eigenvalue $z_a = 0$ of the fifth coalescing eigenstate $|\psi^{z_a}\ket$ exists at all parameter values and is not explicitly shown.
}
\label{fig:EP.surf.1}
\end{figure}

First, let us take a closer look at the expressions for the exceptional points $\gRII$ [Eq. (\ref{EPII})] and $\gRImp$ [Eq. (\ref{EPI})] given in the previous section.  Notice that each of these equations expresses a relationship among the four parameters $t_1, t_2, g$ and $\gamma$.  Since any one of these parameters (say, $t_2$) could be scaled out of the problem, each of these in fact determines a two-dimensional surface of exceptional points in the three-dimensional space defined by the parameters $t_1, g$ and $\gamma$.  Further, recall that $\gRIp$ determines the upper edge of Region I, while $\gRII$ gives the lower edge of Region II.
Reviewing the middle row of Fig. \ref{fig.2}, it is easy to infer that for increasing values of $g$, the upper edge of Region I gradually approaches the lower edge of Region II.  
Hence, to find the exact closing point of the $\PT$ gap we simply set $\gRIp = \gRII$, which yields $g = \ggap$ with
\beq
  \ggap = \frac{t_1 \sqrt{t_1^2 - t_2^2}}{\sqrt{2} t_2}
  	.
\label{g.gap}
\eeq
Plugging this expression back into Eq. (\ref{EPII}) yields the $\gamma$ value at which the gap closes as 
$\gamma = t_1^2/t_2$. 
Notice that whereas there were two free parameters in Eqs. (\ref{EPII}) and (\ref{EPI}), there is only one in Eq. (\ref{g.gap}) [after scaling out $t_2$].  As illustrated in Fig. \ref{fig:EP.surf.1}, this demonstrates that Eq. (\ref{g.gap}) combined with $\gamma = t_1^2/t_2$ defines a curve (purple line) in the $t_1, g, \gamma$ parameter space, which resides precisely at the intersection of the two surfaces $\gamma = \gRIp$ (red surface) and $\gamma = \gRII$ (blue surface) \cite{ENexus}.  The projection of this curve onto the $g, t_1$ parameter space is also shown as the upper red curve in the phase diagram in Fig. \ref{fig.phase}.
Notice that for $t_1 \le t_2$ we find from Eq. (\ref{g.gap}) that the $\ggap$ value vanishes entirely, so that Regions I and II are always directly connected in the topologically trivial phase for $t_1 < t_2$. 

\begin{figure*}[t]
  \includegraphics[width=15cm]{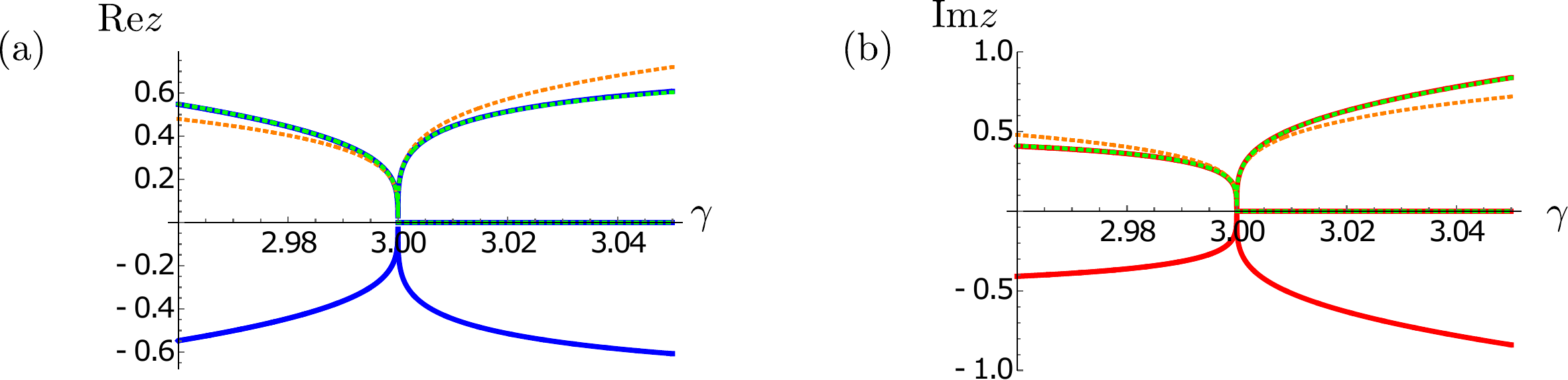}
\caption{ (a) Real part and (b) Imaginary part of the eigenvalue spectrum 
for $t_1 = \sqrt{3}$ and $g = \ggap = \sqrt{3}$ in the unit of $t_2=1$.
The blue solid lines in (a) and the red solid lines in (b) give the exact solutions of eigenvalues.
To give a sense of the increasing accuracy of incorporating additional terms in the Puiseux expansion, the orange dotted line shows only the lowest-order $O (( \gamma^2 - \gEPfp^2 )^{1/4} )$ term from Eq. (\ref{eq.Puiseux1}), while the green dotted line includes the $O (( \gamma^2 - \gEPfp^2 )^{3/4} )$ term.
We here show only the positive values of the expansion solutions for comparison.
}
 \label{fig:spec.EP4}
\end{figure*}

Our analysis below reveals that the exceptional point at the exact gap closing point again involves a coalescence with the localized zero-energy state $|\psi^{z_a}\ket$.  However, in this case, all four of the eigenvalues from the polynomial equation $P_s (z_j) = 0$ are involved.  As a result, the Puiseux expansion for these four eigenvalues involves a quartic root (usually typical of an EP4), while the actual order of the exceptional point appears to be five (EP5).
To illustrate the convergence of the four eigenvalues coming from the polynomial equation, we plot these for the case $t_1 = \sqrt{3}$, $t_2 = 1$ and $g = \ggap = \sqrt{3}$ in Fig. \ref{fig:spec.EP4},  
which clearly demonstrates
the anticipated quartic root behavior in the neighborhood of the gap closing at $\gamma = t_1^2/t_2 = 3$.

In order for each of the four polynomial eigenvalues to coalesce at a single point, all non-trivial derivatives of Eq. (\ref{poly}) must vanish.
The derivatives of the polynomial are given by
\beqa
  P_s' (z) 
    &	= &  4 \gamma^2 z^3 			\label{poly.1}	\\
	& &	+ 2 \left\{ \gamma^4 -2\gamma^2(t_2^2+g^2) + t_1^4 -2g^2t_1^2 \right\} z
					\nonumber 	\\
  P_s'' (z) 
    &	= &  12 \gamma^2 z^2 			\label{poly.2}	\\
        & &	+ 2 \left\{ \gamma^4 -2\gamma^2(t_2^2+g^2) + t_1^4 -2g^2t_1^2 \right\}
					\nonumber 	\\
  P_s''' (z) 
    &	= &  24 \gamma^2 z				\label{poly.3}
    	,
\eeqa
The condition $P_s''' (z) = 0$  from Eq. (\ref{poly.3}) is indeed consistent with $z =  z_a = 0$ at the EP as well as the picture in Fig. \ref{fig:spec.EP4}, which reveals that $\gEPf \neq 0$.  Next, evaluating $P_s'' (z = 0) = 0$ yields
\beq
  \gEPfpm^2
  	= t_2^2 + g^2 \pm \sqrt{ (t_2^2 + g^2)^2 + t_1^2 (2g^2 - t_1^2) } 
	.
\label{gamma.EP4}
\eeq
In the next step, since we already know $z = 0$ at the EP, $P_s' (z=0) = 0$ yields no new information.  So instead, we finally turn back to the original polynomial itself $P_s (z = 0) = 0$ and apply Eq. (\ref{gamma.EP4}) to find
the condition
\beq
  \left( 2 g^2 - t_1^2 + t_2^2 \right)
  	\left[ 2 g^2 t_2^2 - t_1^2 (t_1^2 - t_2^2) \right]
	= 0
	.
\label{EP4.cond}
\eeq
Setting the second factor of this expression to zero yields the same value for $\ggap$ that we previously determined in Eq. (\ref{g.gap}) and confirms the convergence of all four of the polynomial eigenvalues.
Further the closing point of the gap can be shown to occur precisely at $\gamma = \gEPfp = t_1^2/t_2$ (the second sign choice in Eq. (\ref{gamma.EP4}) is spurious in this case, but will turn out to be relevant for the second boundary).  Finally, we can obtain the condition on the wave vector at the EP as $\lambda = - t_2/t_1$, which, comparing with Eq. (\ref{ka.cond}), suggests coalescence with the localized zero-energy state.

To illustrate the behavior of the four polynomial eigenvalues near the EP, we obtain the following Puiseux expansion in the vicinity of the gap closing for the case from Fig. \ref{fig:spec.EP4}, in which $t_1 = \sqrt{3}$, $t_2 = 1$, $g = \ggap = \sqrt{3}$ and, hence, $\gEPfp = 3$.  Applying these parameters in the dispersion polynomial Eq. (\ref{poly}) and expanding in powers of $(\gamma^2 - \gEPfp^2)$, we find
\begin{widetext}
\beq
  z_{1,\{1,2\}} 
  	= \pm \left[ \left( \frac{8}{9} \right)^{1/4} \left( \gamma^2 - \gEPfp^2 \right)^{1/4} 
			- \frac{5}{18} \left( \frac{9}{8} \right)^{1/4} \left( \gamma^2 - \gEPfp^2 \right)^{3/4} 	\right]
			+ O \left(\left( \gamma^2 - \gEPfp^2 \right)^{5/4} \right)		\label{eq.Puiseux1}
\eeq
and
\beq
 z_{1, \{3,4\}} 
  	= \pm i \left[ \left( \frac{8}{9} \right)^{1/4} \left( \gamma^2 - \gEPfp^2 \right)^{1/4} 
			+ \frac{5}{18} \left( \frac{9}{8} \right)^{1/4} \left( \gamma^2 - \gEPfp^2 \right)^{3/4} 	\right]
			+ O \left(\left( \gamma^2 - \gEPfp^2 \right)^{5/4} \right)	,	\label{eq.Puiseux2}
\eeq
\end{widetext}
which are illustrated as the the orange dotted line (lowest-order term only) and green dotted line (both terms) in Fig. \ref{fig:spec.EP4}.

\begin{figure*}[t]
  \includegraphics[width=18cm]{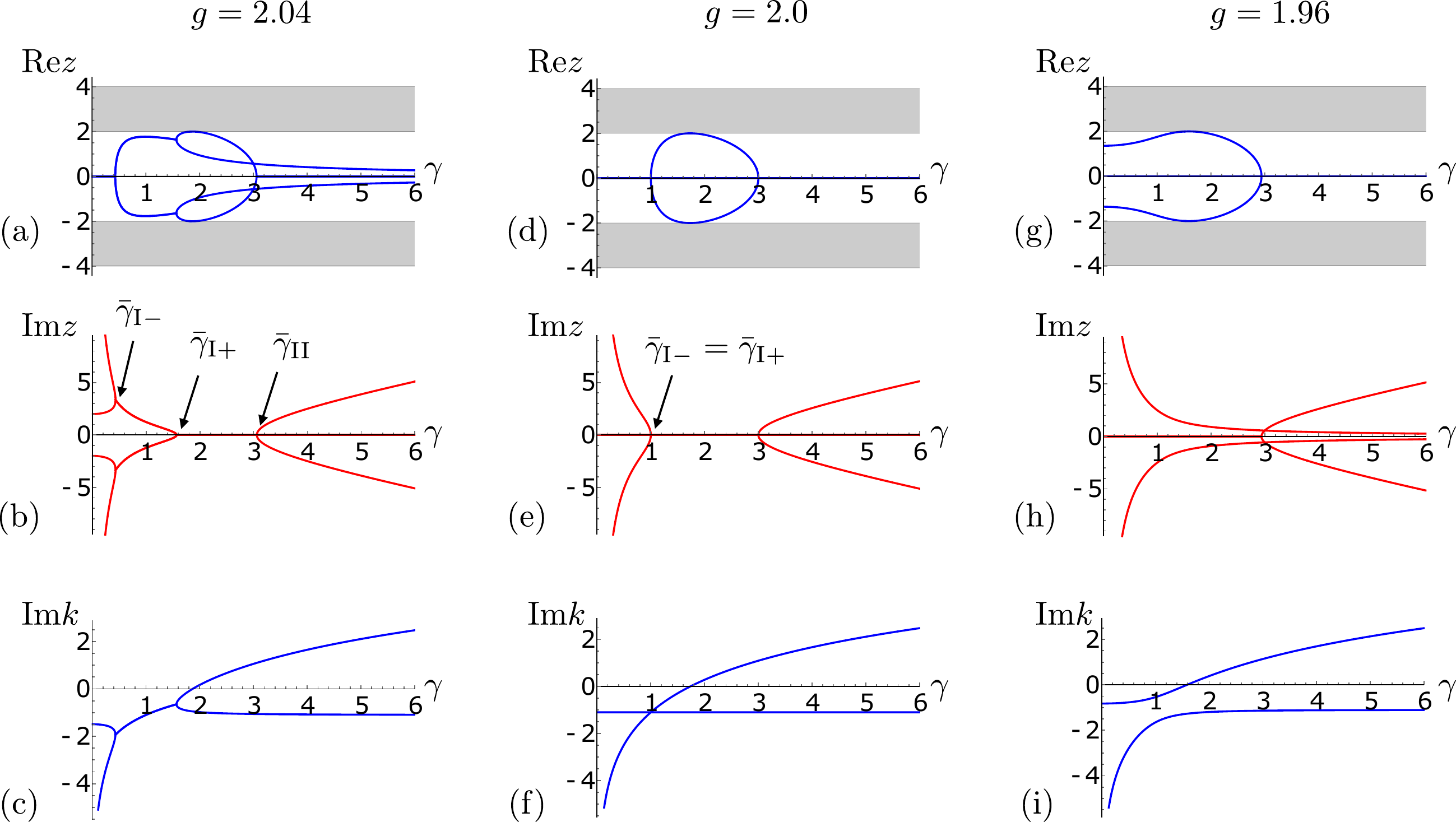}
 \caption{
Discrete eigenvalue spectrum as a function of $\gamma$
for $t_1=3.0$ around $g=g_{\rm gap,2}=2.0$ in the unit of $t_2=1$:
(a) real and (b) imaginary parts of energy and (c) imaginary part of wave number 
for $g=2.04$,
(e) real and (f) imaginary parts of energy and (g) imaginary part of wave number 
for $g=2.0$,
(h) real and (i) imaginary parts of energy and (j) imaginary part of wave number 
for $g=1.96$.
Note that we evaluate the wave number in the unit where the lattice constant is unity.
}
\label{fig.ggap2_1}
\end{figure*}
\begin{figure*}[t]
  \includegraphics[width=18cm]{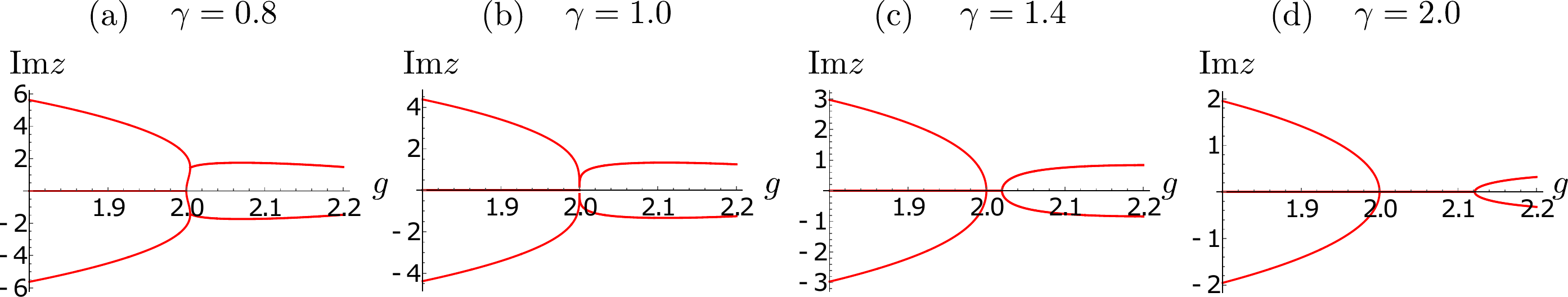}
 \caption{
Discrete eigenvalue spectrum as a function of $g$
for $t_1=3.0$ around $g=g_{\rm gap,2}=2.0$ 
for
(a) $\gamma=0.8$, (b) $\gamma=1.0$, (c) $\gamma=1.4$, and (d) $\gamma=2.0$
in the unit of $t_2=1$.
An EP occurs in each case at $g=2.0$.
}
\label{fig.ggap2_2}
\end{figure*}

In the above discussion, we have determined that the second factor in Eq. (\ref{EP4.cond}) is associated with both the gap closing as well as a coalescence involving one of the zero-energy states.
Hence it is natural to speculate that the first factor in Eq. (\ref{EP4.cond}) might have a similar association.  This turns out to hold true, although the picture is a bit more complicated in this case.  However, before analyzing this issue in detail, it is most natural to first clarify a secondary point.  
Recall that we first noticed the gap closing at $g = \ggap$ by intuiting that the EP2s at the upper edge of Region I and lower edge of Region II 
should connect for sufficiently large $g$.  Similarly, notice that the EP2 $\gRIm $ from Eq. (\ref{EPI}) marking the lower boundary of Region I should eventually vanish as we decrease the value of $g$ (or increase $t_1$).  Indeed, setting $\gRIm = - t_2 + \sqrt{2g^2 + t_2^2 -t_1^2} = 0$ reveals that this occurs for $g = t_1/\sqrt{2}$, which is shown by the dashed blue line in Fig. \ref{fig.phase}.  This means that the {\it lower} $\PT$-symmetric region we encountered in Fig. \ref{fig.2} vanishes for $g \le t_1/\sqrt{2}$.
Hence, below the blue dashed line, for any arbitrarily small value of $\gamma$, the discrete spectrum immediately becomes complex (although within the narrow sliver of parameter space that is shaded red but also falls below the blue line in Fig. \ref{fig.phase}, the $\PT$ gap can still eventually be reached).
We emphasize that below the blue dashed line, the EP2 $\gamma = \gRIm $ still exists, but now the $\PT$-symmetry is broken on either side of this exceptional point and $\gRIm $ should instead be taken as 
$\gRIm = \left| - t_2 + \sqrt{2g^2 + t_2^2 -t_1^2} \right|$.

Now we turn to the first factor 
from Eq. (\ref{EP4.cond}).
Solving this condition $2 g^2 - t_1^2 + t_2^2 = 0$ for $g$ we find $g = \ggapm$ with
\beq
  \ggapm = \sqrt{\frac{t_1^2 - t_2^2}{2}}
  .
\label{g.gap2}
\eeq
This is indicated by the lower red curve in Fig. \ref{fig.phase}, which falls completely below the blue dashed line.
Applying this condition in the dispersion polynomial Eq. (\ref{poly}) we find that the form of all four polynomial solutions dramatically simplifies, even for values of $\gamma$ at which only two (not all four) of the solutions coalesce.  Specifically, for $g = \ggapm$ we find that two eigenvalues become zero and apparently coalesce with the {\it anti-localized} zero-energy state $|\psi^{z_b}\ket$ for all values of $\gamma$, taking the wave vector determined by 
$\lambda = e^{i k_b} = - t_1/t_2$ [in agreement with the condition in Eq. (\ref{kb.cond})].
Then the other two solutions are given by
\beq
  z_{2,\pm} = \pm i \frac{ \sqrt{\left( t_1^2 - \gamma^2 \right) \left( t_2^2 - \gamma^2 \right) } }{\gamma}
  	.
\label{z.EP4.2}
\eeq
As one should immediately suspect, these latter two solutions also coalesce with the zero-energy states for the specific values  $\gamma = \gEPfm = t_2$ and $\gamma = \gEPfp = t_1$.  In the case $\gamma = \gEPfm = t_2$ [corresponding to the minus sign choice in Eq. (\ref{gamma.EP4})] both of $z_{2,\pm}$ seem to coalesce with the other three pre-existing anti-localized states $|\psi^{z_b}\ket$ with eigenvalue $z_b = 0$, forming an EP5.  
Meanwhile for the case $\gamma = \gEPfp = t_1$, these two instead appear to coalesce with the lone localized zero-energy state $|\psi^{z_a}\ket$.  In this last case these appear as two coinciding EP3s with a shared eigenvalue $z_a = z_b =0$.

Let us emphasize a peculiar feature of the $g = \ggapm$ case that is rather different from the previous case $g = \ggap$.  To illustrate the point, in Fig. \ref{fig.ggap2_1} we plot the eigenvalues as we decrease the value of $g$ 
in the vicinity of $\ggapm = 2.0$ for $t_1 = 3.0$ and $t_2 = 1.0$.  
In the left column of Fig. \ref{fig.ggap2_1} the eigenvalues are shown slightly above $\ggapm$ 
at $g = 2.04$, 
in the middle column they are shown exactly 
at $g = \ggapm = 2.0$ 
and in the right column they are shown slightly below 
at $g = 1.96$.  
The key point is that in Fig. \ref{fig.ggap2_1}(d) and (e), there is still an extended gap with pure real eigenvalues that stretches between the two points $\gamma = \gEPfm = 1.0$ to $\gamma = \gEPfp = 3.0$ at which the solutions $z_{2,\pm}$ form EPs.  Further, as previously noted, the two eigenvalues with energy $z=0$ in this case coincide with the the anti-localized zero-energy state.  Hence, the entire gap in this case is actually a line of exceptional points and for any value of $g$ that is infinitesimally smaller than $\ggapm$, the entire gap simultaneously gives rise to complex eigenvalues.
This picture is rather different than the scenario we encountered for $g = \ggap$, in which two EP2s directly connected along with a single zero-energy solution to form the higher-order EP that marked the discrete point at which the gap closed [see Fig. \ref{fig:spec.EP4}].  We further illustrate this difference in Fig. \ref{fig.ggap2_2}, in which we have plotted the imaginary part of the energy eigenvalues instead as a function of $g$ for four different values of $\gamma$ that fall 
below ($\gamma =0.8$), at the edge ($\gamma = 1.0$), and within the gap ($\gamma = 1.4$ and $2.0$).  In each case, an exceptional point clearly marks the expected transition value at $g= \ggapm = 2.0$.

\begin{figure*}
\hspace*{0.05\textwidth}
 \includegraphics[width=0.4\textwidth]{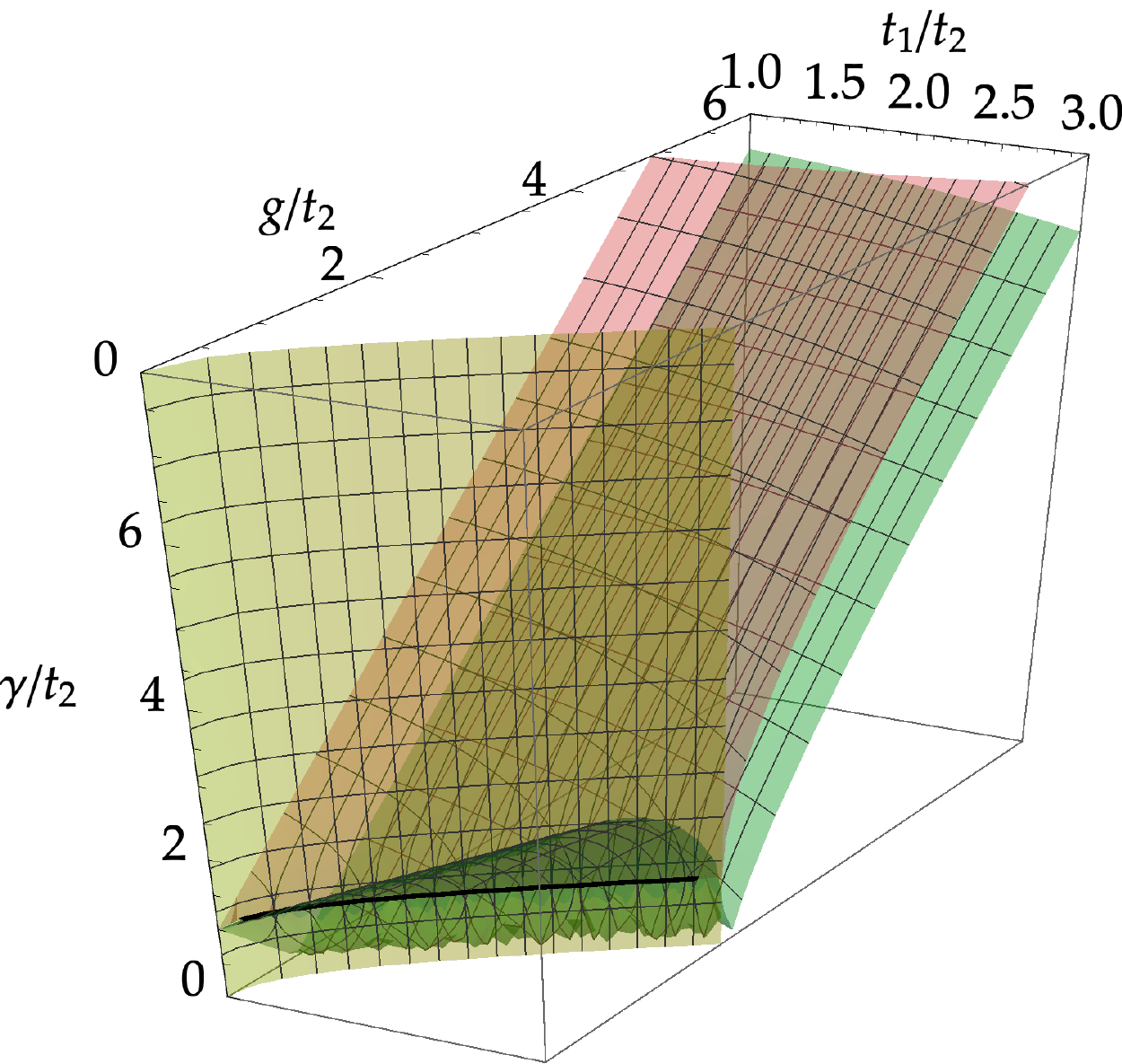}
\hfill
 \includegraphics[width=0.4\textwidth]{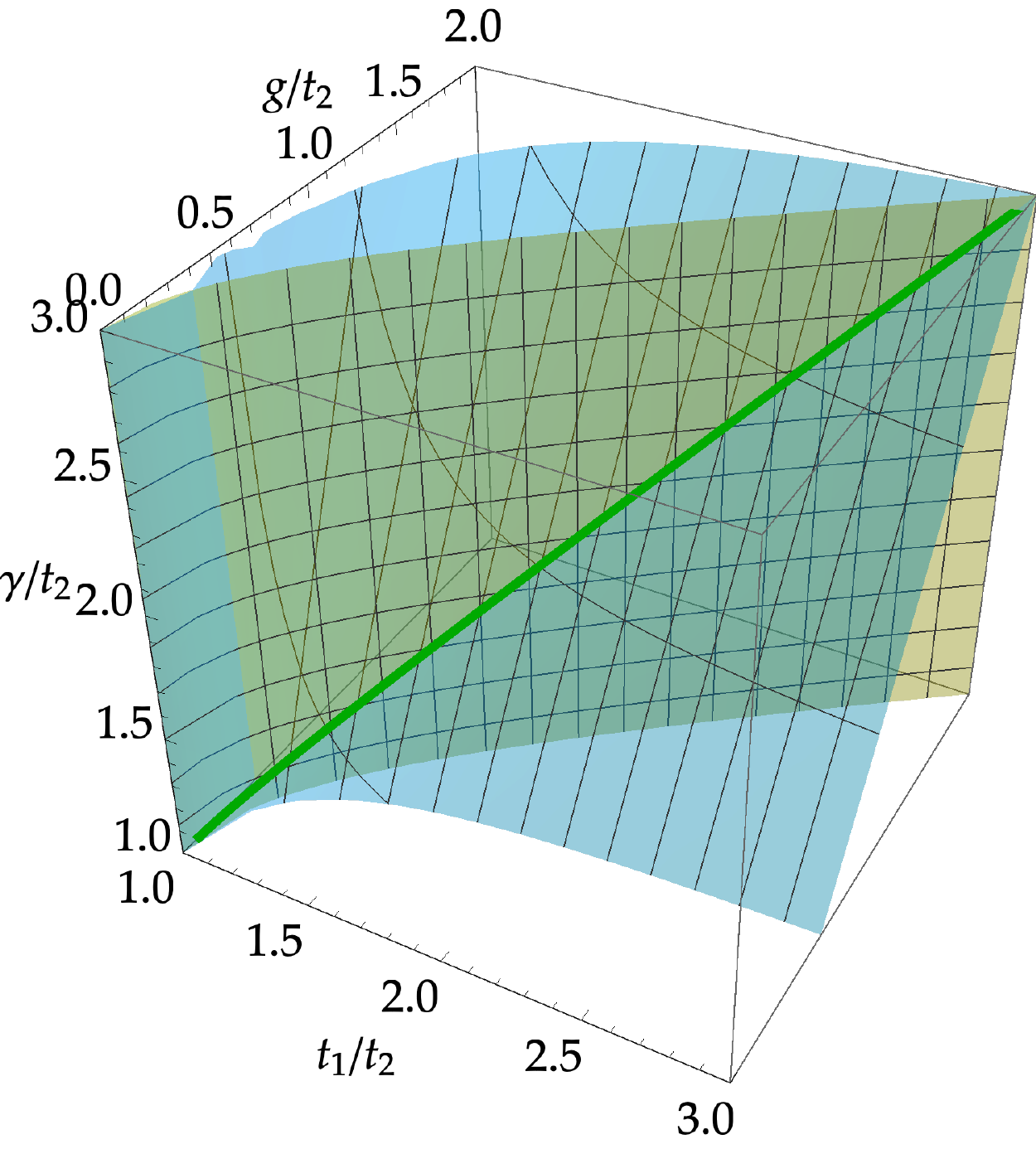}
 \hspace*{0.05\textwidth}
\\
\hspace*{0.01\textwidth}(a)\hspace*{0.47\textwidth}(b)\hspace*{0.4\textwidth}
\\
\caption{
Exceptional surfaces in the 
$(t_1/t_2, g/t_2, \gamma/t_2)$-parametric space 
of the model in the vicinity of the two exceptional curves occurring for $g = \ggapm$.
The EP3 surface occurring along $g = \ggapm$ (with $\gamma \in \left[ 0, \infty \right]$ ) is shown in yellow in both figures.
In (a) the EP2 surface $\gamma = \gRIm = \left| - t_2 + \sqrt{2g^2 + t_2^2 -t_1^2} \right|$  is illustrated in light green while the EP2 surface $\gamma = \gRIp$ is shown in red and the EP5 curve along  $g = \ggapm$ and $\gamma = \gEPfm$ is shown in black (at the intersection of three surfaces).
In (b) the EP2 surface $\gamma = \gRII$ is shown in blue while two coinciding EP3 curves along  $g = \ggapm$ and $\gamma = \gEPfp$ are shown in dark green (at the intersection of two surfaces).  In (b), the sixth coalescing eigenstate with eigenvalue $z_a = 0$ exists at all parameter values and is not explicitly shown.
}
\label{fig:EP.surf} 
\end{figure*}

Hence, the $\PT$-breaking transition in the $g= \ggapm$ occurs spontaneously, simultaneously over an extended range of values of the $\PT$ parameter $\gamma$ as one varies any of the other system parameters.  We have not previously encountered any such $\PT$-transition reported in the literature.  
We also observe from Fig. \ref{fig.ggap2_1}(e) that the EP5 at $\gamma = \gEPfm = t_2$ lies at the exact intersection of {\it three} exceptional surfaces, as shown in the three-dimensional parameter plot in Fig. \ref{fig:EP.surf}(a).
These three surfaces are defined by $\gamma = \gRIm = \left| - t_2 + \sqrt{2g^2 + t_2^2 -t_1^2} \right|$ (EP2 surface, shown in light green), $\gamma = \gRIp = t_2 + \sqrt{2g^2 + t_2^2 -t_1^2} $ (EP2 surface, red) and $g = \ggapm$ (EP3 surface, yellow), while the $\gamma = \gEPfm$ EP5 curve itself is shown in black.
Finally, the $\gamma = \gEPfp$ EP curve (dark green) is shown at the intersection of two EP surfaces $\gamma = \gRII$ (blue, EP2) and  $g = \ggapm$ (yellow, EP3) in Fig.  \ref{fig:EP.surf}(b); however, we emphasize that in this case, the two EP surfaces are separately coalescing with either of the two zero-energy states and hence their shared eigenvalues along the green curve should be interpreted as coincidental.


\section{Dynamics: initial state evolution}\label{sec.surv.prob}

\begin{figure*}
 \includegraphics[width=15cm]{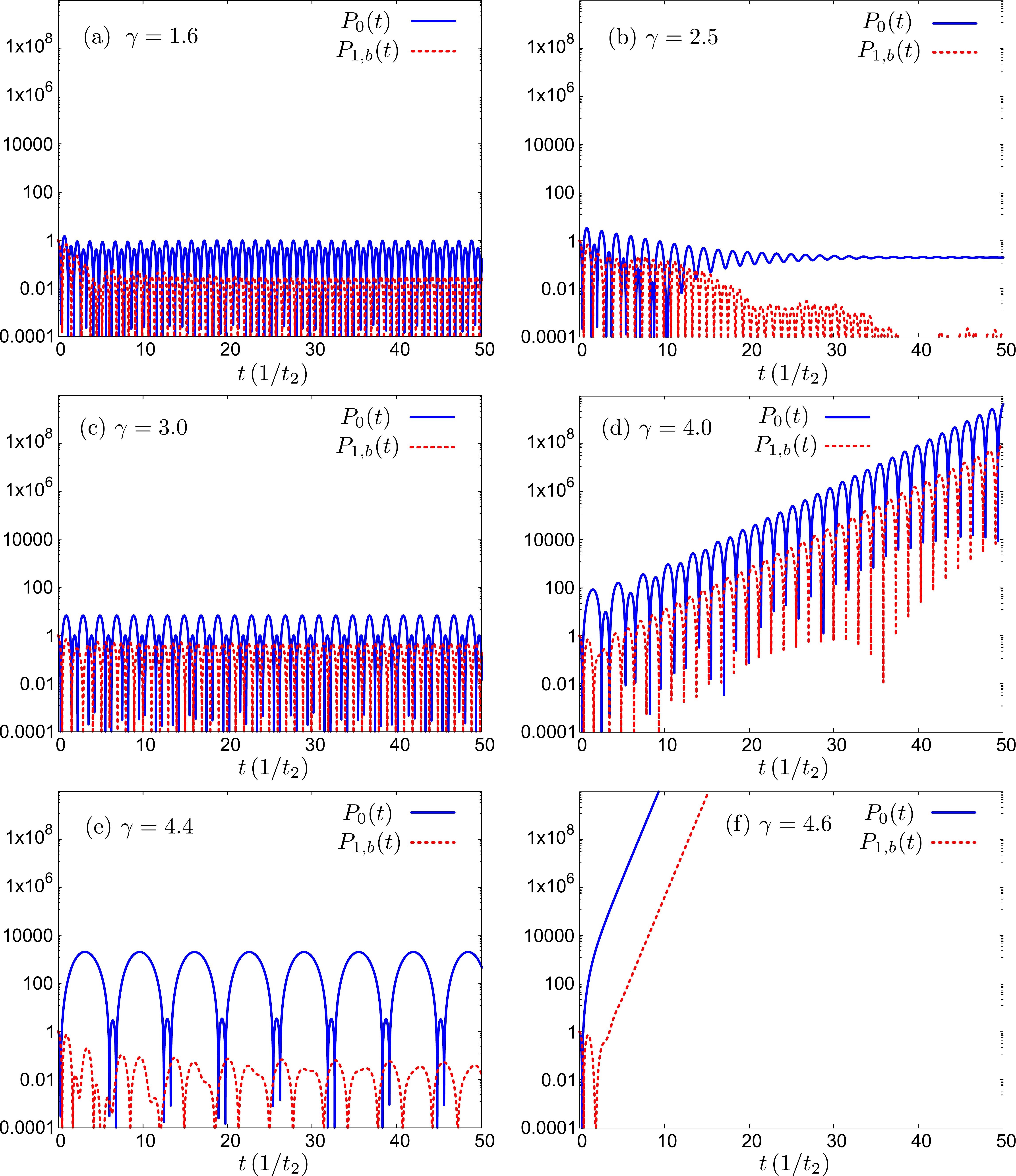}
\caption{
Initial state evolution simulations $P_0(t)$ at the central site $|0\rangle$
(blue solid lines) and $P_{1,b}(t)$ at site $|1,b\rangle$ (red dotted lines)
at time $t$ 
in the case $t_1 =3.0, g=3.0$ 
for (a) $\gamma=1.6$, (b) $\gamma=2.5$,
(c) $\gamma=3.0$, (d) $\gamma =4.0$, (e) $\gamma = 4.4$, and (f) $\gamma=4.6$
in the unit $t_2 = 1$.
The corresponding energy spectrum is shown in Fig.~\ref{fig.2} (d) and (e).
}
\label{fig:gamma} 
\end{figure*}

Several of the results presented in this work could be verified in experiment by observing the evolution dynamics of an initially-prepared state at a given site.  We define the {\it initial state evolution measure} (or initial state measure for short) as follows. 
For the case that the particle is initialized on the central site $| 0 \ket$ this quantity is defined by
\beq
  P_0 (t) \equiv |\bra 0 | e^{-iHt} | 0 \ket|^2
\label{init.meas.0}
\eeq
in terms of the time evolution operator $e^{-iHt}$. 
Physically, this corresponds to the likelihood of finding a particle at site $| 0 \ket$ at time $t$, given the assumption that the initial particle is located at site $| 0 \ket$ such that $P_0 (0) = 1$ while all other sites are empty.
 Similarly, the initial state measure for any arbitrary initialized site $| n, x \ket$ within the SSH leads is written
\beq
  P_{n,x} (t) \equiv |\bra n, x | e^{-iHt} | n, x \ket|^2
\label{init.meas.nx}
\eeq
Note that these quantities might surpass unity for $t > 0$ because additional particles can flow in to the lattice from the gain site in our non-Hermitian system; this is in contrast to Hermitian systems for which the corresponding quantity, the survival probability, always takes values between 0 and 1 (of course, the initial state measure reduces to the survival probability in the Hermitian limit $\gamma \to 0$). 
The most obvious setting in which to observe the initial state measure in the present context would be in a photonic lattice array experiment, similar to Refs. \cite{expt1,expt2}, in which the evolution of an initially-occupied site in our model could be simulated as propagation distance along a given waveguide in the array.  Loss can then be introduced on any waveguide in the array through periodic bending along its length, which enables one to simulate `passive' $\PT$ symmetry \cite{expt1,expt2}. 

In Figs.~\ref{fig:gamma} we present numerical simulations for six values of $\gamma$ corresponding to the $g = 3.0$ case from Fig. \ref{fig.2}, in which every region of the spectrum can be represented.  We show the simulation for two cases in each figure: we present one evolution $P_0 (t)$ for when the particle is initially located at $| 0 \ket$,
and a second $P_{1,b} (t)$ for when it is initialized at site 
$| 1, b \ket$.
Note that the initial state measure for sites $| \pm1, a \ket$ 
(not shown) are qualitatively similar to that for $| 0 \ket$ for these cases, while $|-1, b \ket$ (not shown) would be similar to that for $| 1, b \ket$. These simulations capture the essential features of the eigenstates with respect to $\PT$-symmetry in each region, including the reservoir-assisted symmetry breaking (and the gap separating it from Region II).

First, 
in the cases $\gamma = 1.6$ and 
$\gamma = 2.5$,  shown in Fig.~\ref{fig:gamma} (a) and (b), 
the evolution is qualitatively similar to a traditional Hermitian open quantum system, though not strictly unitary.
The case $\gamma = 1.6$ [Fig.~\ref{fig:gamma} (a)]
corresponds to 
the lowest $\gamma$ region of unbroken $\PT$-symmetry in Fig. \ref{fig.2}(e).  In this case, the initial state measure $P_0 (t)$ is largely driven by Rabi oscillations involving the two ordinary bound states that exist in this case as well as the localized zero-energy state $|\psi^{z_a}\ket$, giving rise to a simple beat pattern. 
Meanwhile for $P_{1,b} (t)$, the Rabi oscillations (appearing after a brief transitory period) exhibit only one oscillation period because the zero-energy state is not involved in this case (recall $|\psi^{z_a}\ket$
has no support on the $b$ sites).  

Next in Fig.~\ref{fig:gamma} (b) 
we present the case $\gamma = 2.5$, 
representative of Region IA shown in Fig. \ref{fig.2}(e).  Here the $\PT$-symmetry is broken while the wave function of all states are anti-localized, which leads to an evolution somewhat similar to traditional resonance decay in the $P_{0} (t)$ simulation, but with fractional decay due to the presence of the localized zero-energy state $|\psi^{z_a}\ket$.
Taking into account the presence of two resonance states with the same imaginary part of the eigenvalue as well as the localized zero-energy state, the evolution can be approximated as $P(t) \sim 1 + 4 D \cos (E_\textrm{R} t + \theta) e^{-\Gamma t/2} + 4 D^2 \cos^2 (E_\textrm{R} t + \theta) e^{-\Gamma t}$  with $E_\textrm{R} \sim 3.5$ the real part of the resonance eigenvalue, $\Gamma \sim 0.25$ the resonance decay width and $D \sim 1$ a constant.
(with the specific approximate values corresponding to Fig.~\ref{fig:gamma} (b) around $\gamma = 2.5$). 
After about $t \sim 30$, 
the evolution settles down to a fractional occupation of the $0$ site due to the localized zero-energy state.  
Meanwhile for the $P_{1,b} (t)$ evolution the dynamics more prominently incorporates non-Markovian decay dynamics associated with the band edges (branch-point effect) \cite{GNOS19,Muga_review,GPSS13,CrespiExpt}
but the most important difference is that the decay is nearly complete because the localized zero-energy state plays no role in this case.

The dynamics at the RIC is shown 
in Fig.~\ref{fig:gamma} (c) for $\gamma = 3.0$, 
in which case the evolution is a simple, almost exact oscillation reflecting the non-equilibrium steady state obtained at the RIC.  Next we turn to the dynamics in Region IB at $\gamma = 4.0$ in Fig.~\ref{fig:gamma} (d).  In this case, the $\PT$-symmetry is broken but the eigenstates are now localized around the non-Hermitian central potential, which leads to the anti-resonance dominating the dynamics and, hence, qualitatively similar exponential growth for both $| 0 \ket$ and $| 1,b \ket$ simulations with growth rate $\Gamma \sim 0.4$.

A representative case for the dynamics in the $\PT$ gap is shown in Fig.~\ref{fig:gamma} (e), in which the $P_{0} (t)$ simulation again shows a relatively simple evolution involving Rabi oscillations with two bound states and the localized zero-energy state.  However, the $P_{1,b} (t)$ dynamics are a bit more complicated with the band edges playing a more prominent role both in the (more complicated) Rabi oscillations and in the gentle, overlying non-Markovian decay pattern.  Finally, for Region II in Fig.~\ref{fig:gamma} (f), both components exhibit the expected exponential growth associated with the highly-localized, dominant anti-resonance with growth rate about $\Gamma \sim 2$.

\begin{figure*}[t]
  \includegraphics[width=8cm]{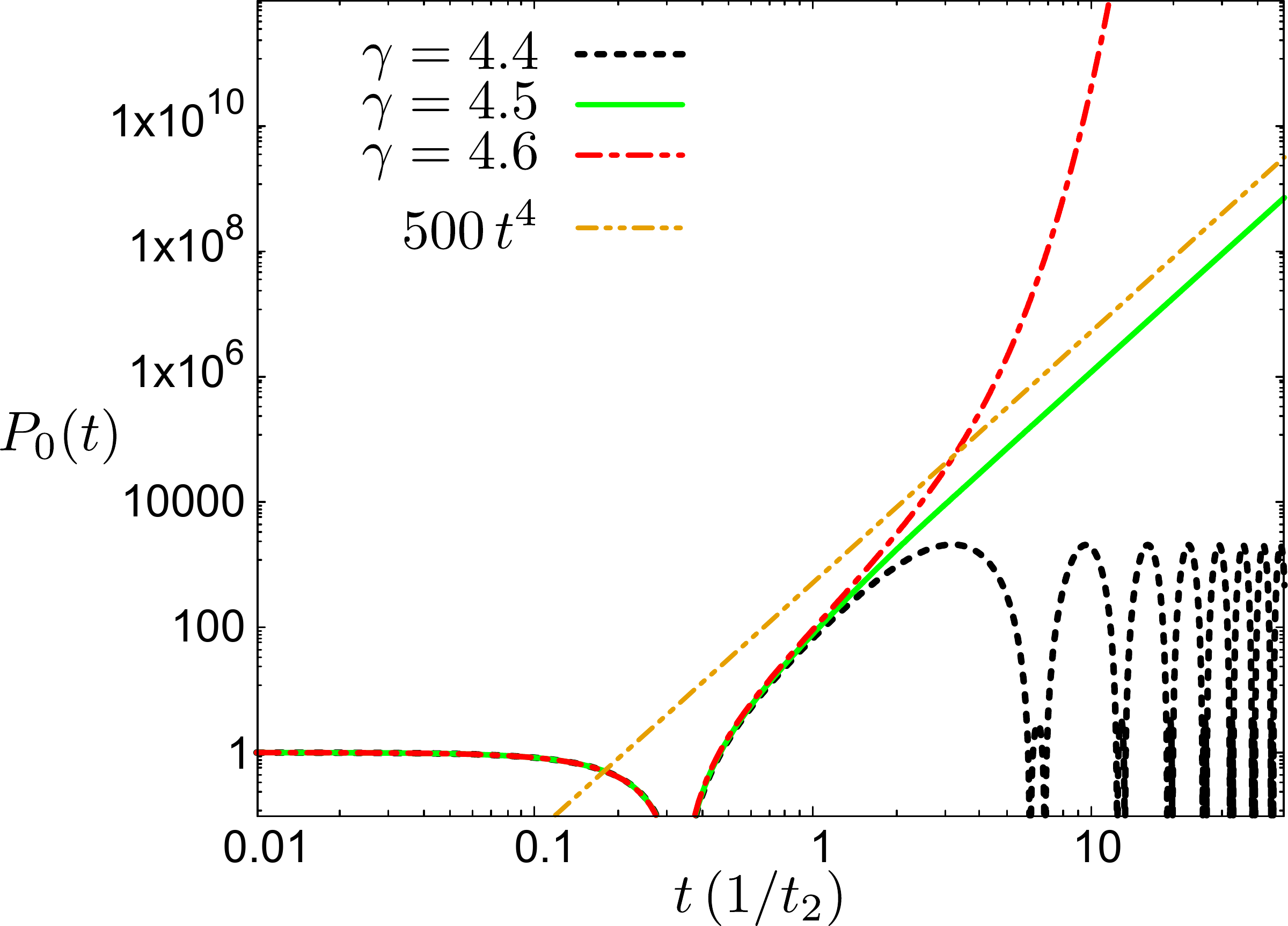}
 \caption{
Initial state measure at the central site $P_0(t)$ 
in case of $t_1 =3.0, g=3.0$ in the unit $t_2 = 1$
for $\gamma=4.4$ in the gap (black dotted line), $\gamma=4.5$ at the threshold
between the gap and Region II (green solid line), 
and $\gamma=4.6$ in Region II (red dot-dashed line).
The corresponding energy spectrum is shown in Fig.~\ref{fig.2} (d) and (e).
}
\label{gamII.EP3}
\end{figure*}

The above outlines an experiment that could be performed to verify the reservoir-assisted symmetry breaking in a photonic lattice experiment. Measuring the oscillatory dynamics in the lowest $\gamma$ region [Fig. \ref{fig:gamma}(a)] would demonstrate the first unbroken $\PT$-symmetric region.  Then measuring the decay dynamics for Region IA in Fig. \ref{fig:gamma}(b) or the growth dynamics for Region IB in Fig. \ref{fig:gamma}(d) [or both] illustrates the broken $\PT$-symmetry in Region I.  The experimentalist could then illustrate the restoration of $\PT$ symmetry in the gap with the oscillatory dynamics of Fig. \ref{fig:gamma}(e), before finally showing that the $\PT$-symmetry is once again broken with the growth dynamics in Region II [Fig. \ref{fig:gamma}(f)].

While our main focus above was the $\PT$-symmetry breaking, we have incidentally also obtained clear evidence of the localized zero-energy state.  First, this appeared in the Rabi oscillations in the $P_{0} (t)$ simulation in the two $\PT$-symmetric regions.  But, more importantly, in Fig.~\ref{fig:gamma} (b) we observed fractional decay in the $P_{0} (t)$ simulation that did {\it not} appear in the $P_{1,b} (t)$ simulation, which gives a more immediately obvious demonstration of the localized zero-energy state.  Indeed, this is roughly comparable to the 
$\PT$-symmetric topological interface state that has been observed in Ref. \cite{expt1} (see Figs. 6(b) and (d) in particular in that work), although with uniformly distributed non-Hermitian defects along the arms of the lattice in that case.

However, we present a more dramatic (and dynamic) method to detect the zero-energy state at (or very near) one of the higher-order exceptional points as follows.  Let us first focus on the case near the exceptional point $\gamma = \gRII$, which, as we discussed following Eq. (\ref{EPII.puiseux}), acts as an EP3 at which two of the polynomial dispersion solutions converge with the localized zero-energy mode  $|\psi^{z_a}\ket$.  In Fig. \ref{gamII.EP3} we plot (in log-log scale) the initial state measure for the central defect site $| 0 \ket$ in three cases: within the $\PT$ gap (black dotted line), directly at the EP $\gamma = \gRII = 4.5$ (green line), and just within Region II (red dot-dashed line) for the case $t_1 = g=3.0 t_2$ corresponding to Fig.~\ref{fig.2} (d) and (e).  We see that the evolution in the gap is clearly bounded, while inside Region II there is exponential growth, with the EP being at the boundary between these two behaviors.  Further, the long-time evolution at the EP is clearly rather simple: pure polynomial growth.

The origin of the power law dynamics can be easily understood at a qualitative level from the order of the pole that would appear under a Green's function (or similar) analysis at the EP.  It is well established \cite{GW64,BG65,GO17,Moiseyev11,Reboiro20,KBH21} that at an EP of order $N$ (EP$N$) with eigenvalue $z$, the $N$th-order pole gives rise to a term in the amplitude for the dynamics of the form
\beq
  \bra \psi | e^{-iHt} | \psi \ket \sim t^{N-1} e^{-i z t}
\eeq
for an arbitrary state $| \psi \ket$ that overlaps with the coalesced eigenvector.
In the present case, with the zero-energy eigenvalue given by $z_a = 0$, this results in pure power law evolution in the initial state measure with the dominant term given by
\beq
   | \bra 0 | e^{-iHt} | 0 \ket |^2 \sim t^{2N-2}
\label{prob.EPN}
\eeq
for the initial state $|0\ket$.
Hence, the $t^4$ evolution observed in Fig.  \ref{gamII.EP3} strongly suggests the coalesced zero-energy state at $\gamma = \gRII$ indeed acts as an EP3.

\begin{figure*}[t]
  \includegraphics[width=8cm]{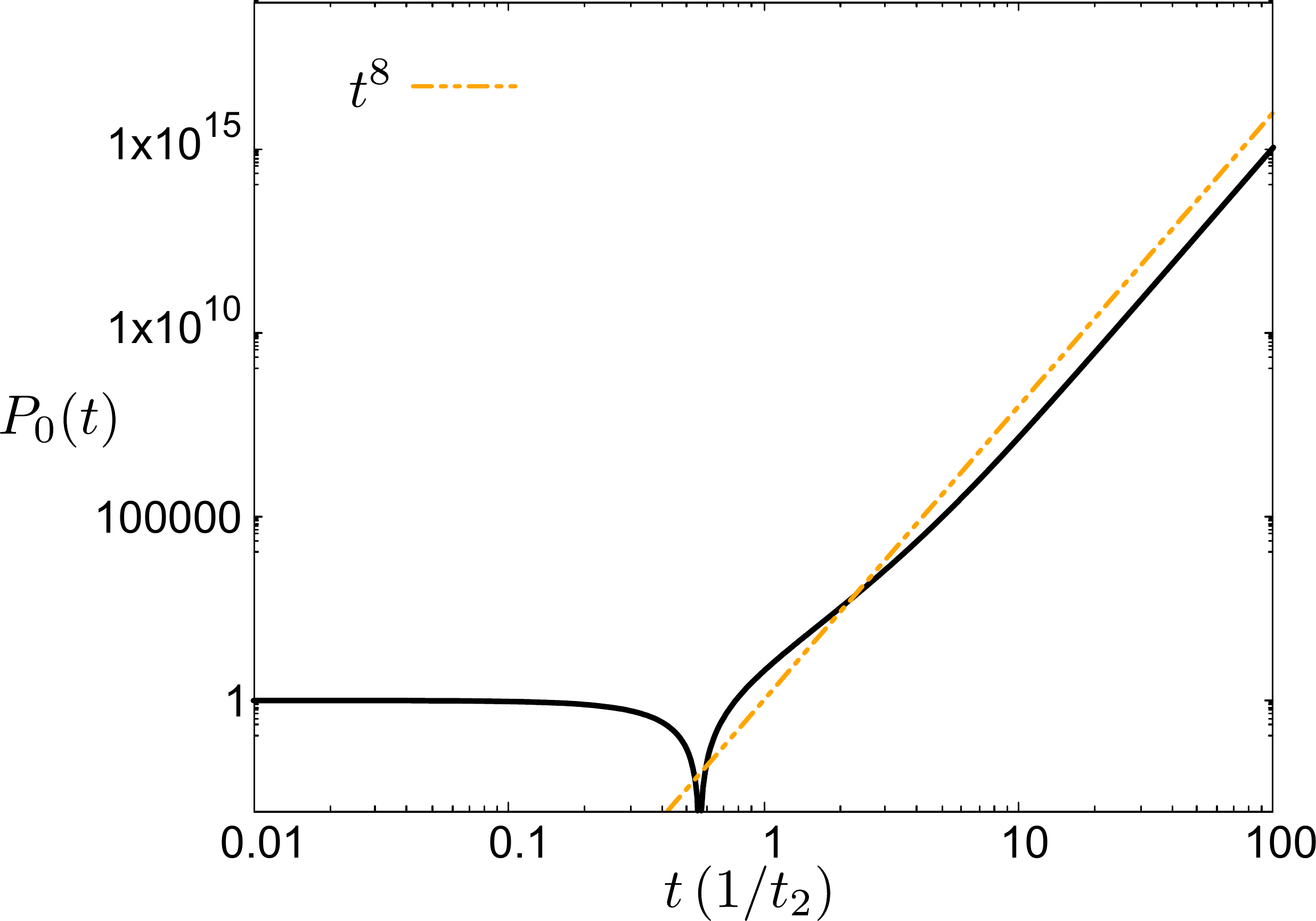}
 \caption{
Initial state measure at the central site $P_0(t)$ 
in the case $t_1 =\sqrt{3}, g=g_{\rm gap,1}=\sqrt{3}$ in the unit $t_2 = 1$
for $\gamma=3.0$ (black solid line). This corresponds to the EP in the spectrum shown in Fig.~\ref{fig:spec.EP4}.
}
\label{ggap.EP5}
\end{figure*}

Next we turn to the dynamics near the zero-energy EP occurring at the upper boundary of the gapped region for $g = \ggap$ at $\gamma = \gEPfp$.  We argued in Sec. \ref{subsec.EP4} that the EP in this case is order five, despite having a Puiseux expansion typical of order four, as was obtained in Eqs. (\ref{eq.Puiseux1}) and (\ref{eq.Puiseux2}) for the case 
$t_1 =\sqrt{3}, g=g_{\rm gap,1}=\sqrt{3}$ and $t_2 = 1$.  For these same parameters, we show the evolution for the initially prepared state $| 0 \ket$ in Fig. \ref{ggap.EP5}, which reveals the dynamics follows $\sim t^8$, which is consistent with an EP5.

\begin{figure*}[t]
  \includegraphics[width=8cm]{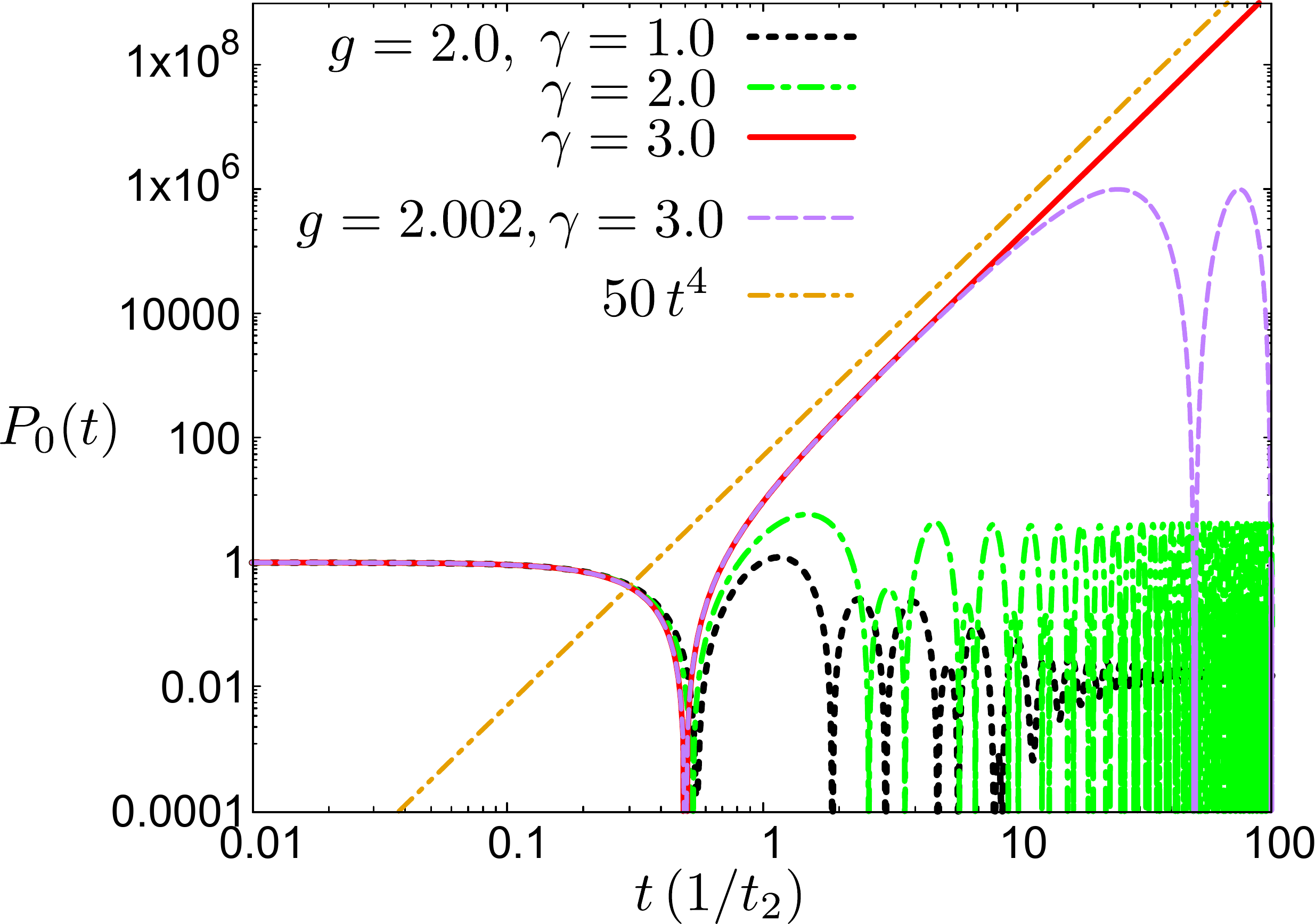}
 \caption{
Initial state measure at the central site $P_0(t)$ 
in case of $t_1 =3.0, g=2.0$ in the unit $t_2 = 1$
for $\gamma=1.0$ (black dotted line), $\gamma=2.0$ (green dot-dashed line),
$\gamma=3.0$ (red solid line).
The corresponding energy spectrum is shown in Fig.~\ref{fig.ggap2_1} (d) and (e).
The simulation for $g=2.002$, $\gamma=3.0, t_1 = 3.0$ is also shown by a purple dashed line.
}
\label{ggap2}
\end{figure*}

Finally, we consider the dynamics in the case $g = \ggapm$ in Fig. \ref{ggap2} for three representative values of $\gamma$, while using the same values $t_1 = 3.0$ and $t_2 = 1$ from Fig. \ref{fig.ggap2_1}.  
For $\gamma = \ggapm = 1.0$ (black dotted line in Fig. \ref{ggap2}) we are exactly at the higher-order exceptional point at which the spectrum consists of 
the lone localized zero-energy state $|\psi^{z_a}\ket$
and an apparent EP5 formed with 
the anti-localized state $|\psi^{z_b}\ket$.  
In this case, we observe clearly non-Markovian decay that settles down to long-time fractional occupation of the localized zero-energy state.  The non-Markovian dynamics in this case can be understood as resulting from the interplay between the anti-localized EP5 and the two nearest (although still rather separated) continuum thresholds, similar to Fig. 4(b) in Ref. \cite{GO17}.  
For the case $\gamma = 2.0$ (green dot-dashed line in Fig. \ref{ggap2}) 
the spectrum consists of two ordinary bound states with eigenvalue $z_{2,\pm}$ from Eq.~(\ref{z.EP4.2}), the localized zero-energy state, and the anti-localized zero-energy EP3.  The dynamics in this 
case are dominated by Rabi oscillations among the bound states, similar to the $\PT$ gap in Fig.  \ref{gamII.EP3}.  Finally, in the case $\gamma = \ggap = 3.0$ (red solid line in Fig. \ref{ggap2}), we see the dynamics again follows a $P_{0} (t) \sim t^4$ evolution, owing to the apparent EP3 formed by 
the coalesced localized state $|\psi^{z_a}\ket$.
Unsurprisingly, the influence from the coalesced anti-localized state 
$|\psi^{z_b}\ket$, 
which is also an EP3 in this case, is completely washed out in the dynamics compared to the localized state.

In Sec. \ref{subsec.zero}, we previously noted that the zero-energy modes satisfy technically different boundary conditions than those satisfied in general by the four polynomial modes.  As a quick consistency check regarding this point, we also include in Fig. \ref{ggap2} a numerical simulation for the dynamics in the case $g=2.002, \gamma = 3.0$ (purple dashed line), just a little away from the highly-singular point at which the two zero-energy EP3s occur at exactly 
$g=2.0, \gamma = 3.0$ 
(and just inside the $\PT$ gap).  We see that the expected $t^4$ evolution is still the most prominent feature in the dynamics, although the evolution is ultimately bounded
since the $\PT$-symmetry is unbroken in this case.

While we have shown here that the influence of the exceptional point formed by the localized zero-energy state should be easily discernible by the satisfyingly simple prediction for the dynamics reported in Eq. (\ref{prob.EPN}), the dynamical influence from
the anti-localized zero-energy state $|\psi^{z_b}\ket$ is much more subtle.  
We discuss this point further in terms of future work below.

\section{Concluding remarks}\label{sec.conclude}

In this paper we have studied $\PT$-symmetry breaking in an open quantum system
consisting of two semi-infinite SSH chains coupled on either side to a $\PT$-symmetric central potential or defect region.
We have characterized the occurrence of two types of $\PT$-symmetry breaking in this system, one of which is primarily induced by the energy continua from the SSH chains and the other that more directly results from the presence of the $\PT$-symmetric potential itself.  We labeled the former as reservoir-assisted $\PT$-symmetry breaking, which occurs for smaller values of the non-Hermitian $\PT$ parameter $\gamma$ than the latter.  Within a significant portion of the parameter space of the model these two types of symmetry breaking were separated by an extended $\PT$ gap in which the reality of the eigenvalues was restored, rendering it easier to distinguish the two types of symmetry-breaking.  The appearance of this gap can be understood as resulting from the energy eigenvalues falling in between the two SSH channels (so the resonance condition is no longer fulfilled) combined with the tendency of the SSH chains to form localized edge states such that the $\PT$-symmetric defect region approximately decouples from the SSH chains.
We finally proposed in Sec. \ref{sec.surv.prob} an experiment to observe the reservoir-assisted $\PT$-symmetry breaking via the initial state evolution dynamics measured in the different regions of the spectrum.

We further pointed out the existence of two zero-energy eigenstates in the spectrum, one of which was localized with respect to the SSH leads and the other of which was anti-localized.  The localized zero-energy state appears to roughly correspond with the $\PT$-symmetric topological interface state observed in the photonic lattice experiment in Ref. \cite{expt1}.  We note that the energy eigenvalue $z=0$ of both the localized and anti-localized state appears in the energy gap directly in the middle of the two SSH energy bands.  Hence, these could more naturally be called {\it mid-gap states}; however, we have not used that term in this paper only to avoid potential confusion with the separate concept of the $\PT$ gap, which is more strictly a property of the discrete spectrum.

Several scenarios occurred in which the closing of the $\PT$ gap gave rise to higher-order exceptional points involving an apparent coalescence of pairs of discrete eigenvalues with either of the two zero-energy eigenstates. Interestingly, in this scenario we found that the Puiseux expansion of the eigenvalues near the EP seems to be one order lower than the actual EP itself.  We compare this with the EP3 with a $\sim k^{1/2}$ dispersion in the non-Hermitian model possessing chiral symmetry studied in Ref. \cite{Mandal21} (see also Ref. \cite{GraefeEP3} for a discussion of this scenario in a more general context).  In our model, for example, the EP marking the closing point of the gap on the upper boundary of the phase diagram in Fig. \ref{fig.phase} has a Puiseux expansion typical of an EP4 \cite{JoglekarEP4,Ghosh19,CBS21,ZnojilEP4}, although we found it behaves like an EP5 \cite{ZnojilEP4}.  In that case, the EP is formed with the localized zero-energy state $|\psi^{z_a}\ket$.  
We further discussed that exceptional points involving the localized zero-energy state should result in characteristic power-law dynamics of the form $P(t) \sim t^{2N -2}$, in which $N$ is the order of the EP.  This could provide a signature of the localized zero-energy EP that we suggest might be observed in a modified version of the experiment in Ref. \cite{expt1}. 

We have said less in this work about the potential dynamical influence of the anti-localized zero-energy state.  Because these states are, by definition, localized away from the central potential (that is usually of primary physical interest) their influence is necessarily much more challenging to detect.  However, previous work has shown that the influence of anti-localized states can be more directly felt in the survival probability dynamics in the case that they appear very close to one of the band edges in the system \cite{GO17,GNOS19,DBP08,GPSS13}, although this requires rather fine tuning of the system parameters. We leave a closer investigation of this subtle effect to future work. 

We end the paper with a final comment on the reservoir-assisted symmetry breaking.  
As we showed in Sec.~\ref{subsec.RAPTB}, this effect in Region I
can, in general, be divided into two qualitatively different subregions.
The resonance state appearing in Region IA has a spatially divergent wave function, which is 
 qualitatively similar to the resonance appearing in ordinary 
 Hermitian open quantum systems \cite{PPT91,HSNP08,GNHP09,DBP08}. This is consistent with the dynamical evolution we obtained for Region IA in Sec. \ref{sec.surv.prob}, although we observed that the dynamics can exhibit non-unitary effects in the present case.
Meanwhile in Region IB, even the similarity for the resonance with the Hermitian picture breaks down as the wave function for the states with complex eigenvalue now becomes localized.  But since the real part of the eigenvalues for these states resides within the continuum, they have been interpreted in Refs. \cite{GGH15,ZK20,LonghiQBIC,KMKT18} as representing a quasi-bound state in continuum.  This interpretation is consistent with the experimental study of defect states in a $\PT$-symmetric optical lattice in Ref. \cite{PTdefectExpt}.

\section*{Acknowledgments}
The authors thank F. Roccati for helpful comments on a previous draft.
We also thank the referee for helpful comments that improved the manuscript.
We also thank
A. Fring,
N. Hatano,
K. Imura,
K. Kanki,
T. Taira,
and S. Tanaka
for fruitful discussions. 
This work was supported by the Japan Society for the Promotion of Science under KAKENHI Grant No. JP18K03466.



\end{document}